\documentclass[twocolumn,amssymb,10pt,prd,nofootinbib,preprintnumbers]{revtex4}
\usepackage{epsfig}

\newcommand{\be}{\begin{equation}}
\newcommand{\ee}{\end{equation}}
\newcommand{\bea}{\begin{eqnarray}}
\newcommand{\eea}{\end{eqnarray}}

\newcommand{\non}{\nonumber \\}

\def\half{{1 \over 2}}

\def\pa{{\partial}}

\def\({\left(} \def\){\right)}
\def\[{\left[} \def\]{\right]}
\newcommand{\ltsim}{\protect\raisebox{-0.5ex}{$\:\stackrel{\textstyle <}{\sim}\:$}}
\newcommand{\gtsim}{\protect\raisebox{-0.5ex}{$\:\stackrel{\textstyle >}{\sim}\:$}}

\newcommand{\h}[1]{{\hat #1}}
\newcommand{\tl}[1]{{\tilde #1}}

\def\cC{{\cal C}}
\def\cL{{\cal L}}

\def\R{{\cal R}}
\def\J{{\cal J}}

\def\al{\alpha}
\def\bt{\beta}
\def\gm{\gamma}

\def\lam{\lambda}

\def\eps{\epsilon}

\preprint{AEI-2009-108}

\begin{document}

\title{\center{An axisymmetric generalized harmonic evolution code}}
\author{Evgeny Sorkin} \email{Evgeny.Sorkin@aei.mpg.de}
\affiliation{Max Planck Institute for Gravitational Physics (Albert Einstein Institute)
Am Muehlenberg 1,
D-14476, Golm, Germany}

\begin{abstract}
We describe the first axisymmetric numerical code based on the generalized harmonic formulation of the Einstein
equations, which is regular at the axis. We test the code
by investigating gravitational collapse of distributions of complex scalar field in a Kaluza-Klein spacetime.
One of the key issues of the harmonic formulation is the choice of the gauge
source functions, and we conclude that a damped wave gauge is remarkably robust in this case.
Our preliminary study indicates that evolution of regular initial data leads to formation both of
black holes with spherical and cylindrical horizon topologies.
Intriguingly, we find evidence that near
threshold for black hole formation the number of outcomes proliferates. Specifically,
the collapsing matter splits into individual pulses, two of which travel in
the opposite directions along the compact dimension and one which is ejected radially from the axis.
Depending on the initial conditions, a curvature singularity develops inside the pulses.
\end{abstract}

\maketitle

\section{Introduction}
\label{sec_intro}

In general, a detailed investigation of fully nonlinear gravitational dynamics
is impossible by other than numerical means. Luckily, the numerical methods have recently reached
the level of maturity that finally allows addressing many long-standing puzzles.
Perhaps the most remarkable is the progress achieved in solving a
general-relativistic two-body problem---the coalescence of black holes
\cite{FP2,FP1,FP3,RIT,nasa,AEI,CaltechCornell}. Driven by the
gravitational wave detection prospects, the problem of the
collision of two black holes or neutron stars continues to
be the central front where an overwhelming majority of numerical
relativity research is done. Fortunately, the computational methods employed there are
portable and---as demonstrated below---can readily be applied
on other problems of interest.

The success of the numerical simulations was backed up by a
parallel development of the software and the hardware, which provided
the necessary computational resources. The rapid hardware evolution combined with the
persisting regularity problems in axial symmetry eventually led to
a direct transition from highly symmetrical spherical configurations
to fully general 3D situations without any symmetries at all,
essentially bypassing the intermediate axisymmetric case.
However, here we argue that important theoretical and practical
reasons exist to explore axisymmetry better, and
we describe a new regular numerical code that, we believe, will be capable of
achieving this.~\footnote{See \cite{Garfinkle_axi,Choptuik_axi,Rinne_axi_1,Rinne_axi_2}
  for alternative approaches.}

Before describing any concrete setup we would point out one important possible
use of an axisymmetric code, specifically, that
it can be regarded as an efficient ``calibration tool'' for more
general 3D codes. \footnote{and as a probe of reliability of the
  cartoon methods \cite{cartoon_method,FP1} used to effectively
  simulate axisymmetric spacetimes in 3D.}  Indeed, we expect that an
intrinsically axisymmetric code applied to, say, the head-on collision of
two black holes would be capable of following the evolution and the resulting gravitational radiation
more accurately compared to Cartesian 3D numerical implementations, both because of explicit use of
the symmetry and since higher numerical resolution can be employed
for given hardware resources.

Several interesting open problems arise in axisymmetric gravitational
collapse situations.  In particular, it remains unclear whether or not
the weak cosmic censorship is violated in collapse of prolate Brill
waves \cite{BrillWaves,Rinne_axi_1}. An independent observation of
universality in critical collapse of gravitational waves \cite{AbrahamsEvans}
is pending, as well as further investigation of
the non-spherical unstable mode that apparently shows up
at threshold for black hole formation in axisymmetric collapse
of a scalar field \cite{crit_collapse_axi}. A basic problem of mathematical
relativity concerning the stability of black holes
with respect to nonlinear axisymmetric perturbations, can
be equivalently addressed in a collapse situation by computing how
fast a newly formed black hole radiates away higher multipole moments.

In addition, we shall mention other, rarely cited in the
context of numerical relativity, axisymmetric systems which are of great interest in
theoretical work on higher-dimensional gravity.  One of the most basic
motivations for studying higher-dimensional spacetimes relies on the
observation that the Einstein equations, describing classical General
Relativity (GR), are independent of the spacetime dimension.
Nevertheless, certain properties of the solutions to these equations vary
dramatically with the dimension.  A striking example is that axisymmetric black
holes in dimensions greater than four do not necessarily have
spherical horizons, but also admit horizons of toroidal topology
\cite{BlackRing}. Moreover, and in sharp contrast to
their four-dimensional counterparts, higher dimensional black holes do not
respect the Kerr limit \cite{MP}, and they are unstable in
certain range of parameters \cite{GL}. One of the fundamental unresolved
puzzles \cite{HM,Choptuik_BS} is whether or not the instability leads
to fragmentation of the horizon and exposition of the inner singularity,
hence violating the cosmic censorship hypothesis.

Since it is unlikely that any of these these problems can be addressed
analytically in a systematic manner, we turn to computational
methods.  However, solving the Einstein equations numerically is
notoriously difficult and depends crucially on the way
these equations are formulated and evolved.  In this paper we focus on
the generalized harmonic (GH) formulation
\cite{Friedrich85,Friedrich96,Garfinkle2001} that has recently gained
popularity because of its great success in the simulations of black
hole binaries \cite{FP2,FP3,CaltechCornell,FP1,Lindblom_etal}.

In a nutshell, the GH approach is a way to write the field equations
such that the resulting system is manifestly hyperbolic, taking the
form of a set of quasilinear wave equations for the metric
components. As the name suggests, the GH method generalizes the
harmonic approach, which achieves strong hyperbolicity by choosing the
harmonic spacetime coordinates. In the GH approach much of the
coordinate freedom is regained through the introduction of certain
gauge source functions, which also maintains the
desirable property of strong hyperbolicity of the field equations. In
fact, the source functions can be thought of as representing the
coordinate freedom of the Einstein equations, and when constructing
solutions of the equations, via an initial value approach, for
example, they must be completely specified in some fashion. Choosing the source
functions in a controlled way is a key issue of the GH technique and after testing several
recent prescription, we conclude that a damped wave gauge \cite{LS}
is remarkably robust in collapse situations.

Applications of the GH approach in spherically-symmetric situations
were studied in \cite{SorkinChoptuik} and here we employ the method in
axisymmetry. In both cases it is natural to use coordinates in which
the symmetries of the spacetime are explicit. However, these
coordinates, are formally singular: at the origin in spherical
symmetry and on the axis in axial symmetry.  Thus, the field
equations have to be regularized in numerical implementations;
here we describe a regularization procedure that is compatible with
the GH formulation.

Our numerical implementation of the GH system is a free evolution
code that advances initial data by solving a set of wave equations.
In addition, there are also constraint equations that must be
satisfied during the evolution.  Although the constraints are
consistently preserved in the GH approach in the continuum limit, in
numerical computations at finite resolution, constraint violations
generically develop.  In order to maintain stability these deviations
must be damped and we discuss an effective method that achieves this.

Having in mind implications for higher-dimensional GR we
test our new code by studying gravitational
collapse of a complex scalar field
in a $D$-dimensional Kaluza-Klein (KK) spacetime.  This background,
which has a single compact extra dimension curled into a circle, is a
classical example of a higher-dimensional compactified spacetime that
in certain limits can appear four dimensional, for example when the
size of the compact dimension is small.\footnote{In this paper,
  however, the size of the KK circle is arbitrary.}
Assuming spherical symmetry in the infinite $(D-2)$-dimensional portion of the space
makes the problem 2+1 that depends on time and two spatial
coordinates: one in the radial direction, and one along the KK
circle.

We perform a series of numerical simulations where the initial
distribution of the scalar matter is freely specified and the outcome
of the evolution depends on the ``strength'' of the initial data as
well as on its topology.  The weak data correspond to the dispersion
of relatively dilute pulses, while a typical strong data configuration
leads to black hole formation or nearly does so. Our preliminary
results indicate a wide range of the black hole forming scenarios,
including how many holes form and of what topology.  For
instance, a static distribution of matter centered at the axis and
localized in the KK direction with the energy-density above certain
threshold collapses to form a black hole with a quasi-cylindrical horizon,
smeared along the extra-dimension.  Data with the energy-density below
this threshold evolve to form a quasi-spherical horizon
centered around the initial matter distribution.
By further reducing the density we find that resulting black holes become
progressively smaller, and at some critical density a pulse of matter is emitted
radially away from the axis and a curvature singularity develops inside it.
We find that for slightly lower initial densities
the evolving matter splits into several pulses and two of them
individually collapse to form the singularities moving apart along the KK dimension.
Finally, when the initial matter distribution becomes dilute below a certain limit,
no black holes or curvature singularities are created.

In the next section we describe the class of effectively 2+1
dimensional models where our code can currently be applied. In Sec.
\ref{sec_GH} we present the basic formulas of the GH formulation and discuss
the constraints and a method to damp their violations.
We describe an axis regularization in Sec. \ref{sec_axis} and boundary conditions  in Sec. \ref{sec_eqs}.
Although, in this work we integrate full $D$-dimensional equations, in
the Appendix we describe an alternative approach---one that uses a dimensional
reduction on the symmetry and integrates the reduced 2+1
equations---and compare its performance with ours.  Coordinate
conditions and the initial data problem are formulated in Secs.
\ref{sec_coord_condition} and \ref{sec_id}, respectively. We use
several diagnostics to probe the spacetimes that we construct,
including computation of asymptotic measurables and apparent horizons,
and describe that in Sec. \ref{sec_diagnostics}.  After elaborating on
our numerical algorithm in Sec. \ref{sec_numercal_approach} we test
its performance in Sec. \ref{sec_num_experements}, giving detailed
accounts of various aspects, such as specific coordinate choices,
constraint damping, numerical dissipation, and convergence.
We conclude in Sec. \ref{sec_conclusion}, outline
possibilities of improvement, and discuss some future prospects.

\section{The Setup}
\label{sec_setup}
We consider a $D$-dimensional spacetime that possesses the $O(D-2)$ isometry group. We will further assume
that the corresponding Killing vectors are orthogonal to the closed
hypersurface they generate.  The symmetry reduces the problem
to effectively 2+1, which depends on time, $t$, and two spatial
dimensions that we denote by $r$ and $z$. The spatial coordinates can
be either infinite or finite.  For instance, taking $D=4$ and assuming
asymptotic flatness correspond to the usual four-dimensional
axially-symmetric situation without angular momentum, while setting
$D=5$ and assuming periodic $z$ and infinite $r$ can describe dynamics
in a five-dimensional Kaluza-Klein background.

The most general $D$-dimensional line element with these isometries
can be written as
\bea
\label{metric}
ds^2 &=& g_{\mu\nu}\, dx^\mu dx^\nu =\non &=& g_{ab}\, dx^a dx^b +
e^{2\, S}\,r^2\, d\Omega_n^2.  \eea
Here $g_{\mu\nu}$ is the $D$-dimensional metric, $ d\Omega_{n}^2$ is
the metric on a unit n-sphere, $n\equiv D-3$, $a,b = 0,1,2$ running
over $\{t,r,z\}$, and the $3$-metric $g_{a b}$ and scalar $S$ are
functions of $t,r$ and $z$, alone.

We take a complex, minimally coupled to gravity scalar field to
represent the matter of the theory and write the total action of
the system as
\bea
\label{action}
S&=&S_{EH}+S_\Phi=\non &=&\frac{1}{16\pi G_N} \int \sqrt{-g_D} \Big[
R_D- \non &-&g^{\mu\nu}\,\pa_{(\mu}\Phi\,\pa_{\nu)}\Phi^* +
2\,V(|\Phi|) \Big] dx^D, \eea
where $G_N$ is the $D$-dimensional Newton constant.

In the next section we will describe our strategy to solving the
equations derived from this action.  We will focus on asymptotically
flat spacetimes times a circle (having the topology
$\mathbb{R}^{D-2,1}\times S^1$) but remark that asymptotically de
Sitter (dS) and anti-de Sitter (AdS) spacetime are also included in
our model (\ref{action}) when the potential of the scalar field
satisfies $V(0)\rightarrow \Lambda$ with positive and negative
$\Lambda$, respectively.

\section{Generalized-harmonic formulation and its reduction to 2D
  case}
\label{sec_GH}
In order to numerically solve the Einstein equations derived from
(\ref{action}), we use the generalized harmonic formulation.  To make
the description as self-contained as possible, we summarize below basic
facts regarding the approach, (more details can be found in e.g.
\cite{FP1,Lindblom_etal,SorkinChoptuik}) and adapt it to the 2+1
situation of interest.

We begin by noting that whenever isometries are present one could
perform a Kaluza-Klein reduction on them, and in our case
(\ref{action}) the reduction yields lower-dimensional, 2+1 Einstein
equations coupled to the scalar, which is related to the size of the
$n$-sphere, and the matter.  Initially, we had indeed performed such
a reduction and coded the reduced equations; see the Appendix \ref{sec_KK}
for details and comparison of the methods. However, after
experimenting with the reduced and the full $D$-dimensional versions
of the equations, we found that numerical solution of the latter is
generically more stable.  Therefore, in what follows we adopt the
unreduced approach.

The Einstein equations on a $D$-dimensional spacetime obtained by
varying the action (\ref{action}) can be written in the form
\be
\label{Eeqs}
R_{\mu\nu}=8 \pi G_N \bar{T}_{\mu\nu}\equiv 8 \pi G_N
\(T_{\mu\nu}-\frac{1}{D-2} g_{\mu\nu} T \), \ee
where $R_{\mu\nu}$ is the Ricci tensor,
$T_{\mu\nu}=\pa_{(\mu}\Phi\,\pa_{\nu)}\Phi^*-(1/2) g_{\mu\nu}
(|\pa\Phi|^2+2\,V )$ is the energy-momentum tensor of the matter with
trace $T$. Hereafter we use units in which the $D$-dimensional Newton
constant satisfies $8\pi\,G_N=1$.

The Ricci tensor that appears in the left-hand-side of (\ref{Eeqs})
contains various second derivatives of the metric components
$g_{\mu\nu}$: these second derivatives collectively constitute the
principal part of $R_{\mu\nu}$, viewed as an operator on $g_{\mu\nu}$.
This principal part can be decomposed into a term $g^{\al\bt}
\pa_{\al\bt} g_{\mu\nu}$, plus mixed derivatives of the form $g^{\al
  \gm} \pa_{\al\mu} g_{\gm\nu}$. Without the mixed derivatives,
(\ref{Eeqs}) would represent manifestly (and strongly) hyperbolic wave
equations for the $g_{\mu\nu}$~\cite{Friedrich96}.  One can view the GH
 formulation of general relativity as a
particular method that eliminates the mixed second derivatives
appearing
in~(\ref{Eeqs}); see ~\cite{Friedrich85,Garfinkle2001,FP1,FP3,Lindblom_etal}.

One requires that the coordinates satisfy
\be
\label{GH_coords}
\Box x^\al = -\Gamma^\al= H^\al, \ee
where $H_\al\equiv g_{\alpha\beta}H^\beta$ are arbitrary ``gauge
source functions'' which are to be viewed as specified quantities, and
$-\Gamma^\al \equiv \Gamma^\al_{\mu\nu} g^{\mu\nu}$ are the contracted
Christoffel symbols.  One defines the GH constraint
\be
\label{Ca}
C^\al \equiv H^\al - \Box x^\al, \ee
which clearly must vanish provided~(\ref{GH_coords}) holds, and then
modifies the Einstein equations as follows:
\be
\label{eqH0}
R_{\mu\nu} - C_{(\mu;\nu)} = \bar{T}_{\mu\nu}. \ee
This last equation can be written more explicitly as
\bea \label{eqH} &-&\half g^{\al\bt} g_{\mu\nu,\al\bt} -
{g^{\al\bt}}_{(,\mu} g_{\nu)\bt,\al} - H_{(\mu,\nu)} +
H_\bt\Gamma^\bt_{\mu\nu} -\non &-&\Gamma^\al_{\nu\bt}
\Gamma^\bt_{\mu\al}=\pa_{(a} \Phi \, \pa_{b)} \Phi^* + \frac{2}{D-2}
g_{ab}\, V.  \eea
Now, provided that $H_\al$ are functions of the coordinates and the
metric only, but not of the metric derivatives---namely
$H_\al=H_\al(x^\mu,g)$---the field equations~(\ref{eqH}) form a
manifestly hyperbolic system. The source functions $H_\al$ are
arbitrary at this stage and their specification is equivalent to
choosing the coordinate system (``fixing the gauge'').  Determining an
effective prescription for the source functions is thus crucial for
the efficacy of the GH approach, and our strategies for fixing the
$H_\al$ are discussed in Sec. \ref{sec_coord_condition}.

After the coordinates have been chosen, we integrate the equations forward
in time.  Consistency of the scheme requires that the GH constraint
(\ref{Ca}) be preserved in time.  The contracted Bianchi identities
guarantee that this is indeed the case, since, using those identities,
one can show \cite{FP1,Lindblom_etal} that $C^\al$ itself satisfies a
wave equation,
\be
\label{C_eq}
\Box C^\al +{R^\al}_\nu \, C^\nu =0. \ee
Thus, assuming that the evolution is generated from an initial
hypersurface on which $C^\al=\pa_t C^\al=0$, and constraint preserving boundary
conditions are used during the evolution, (\ref{C_eq}) guarantees
that $C^\al=0$ for all future (or past) times.

Although the GH constraint is preserved at the continuum level, in
numerical calculations, where equations are discretized on a lattice
the constraint cannot be expected to hold exactly.  It appears that
numerical solutions of~(\ref{eqH}) can admit ``constraint violating
modes'', with the result that the desired continuum solution is not
obtained in the limit of vanishing mesh size.  However, an effective
way of preventing the development of such modes in numerical
calculations exists: one adds terms to the field equations that are
explicitly designed to damp constraint violations (see
e.g.~\cite{KST}).  Following the approach of Pretorius \cite{FP1,FP3}
that builds on earlier works \cite{gm_sys,Gundlachetal} we define the
constraint damping terms
\be \label{cdmp}
Z_{\mu\nu} \equiv \kappa \( n_{(\mu}\cC_{\nu)} -\half
g_{\mu\nu} \, n^\bt \, \cC_\bt \), \ee
and solve the modified equations of the form
\be
\label{Eqs_constrdamp}
R_{\mu\nu} - C_{(\mu;\nu)} +Z_{\mu\nu}= \bar{T}_{\mu\nu}.  \ee
Here, $n_\mu$ is the unit time-like vector normal to the $t={\rm
  const.}$ hypersurfaces, that can be written as
\be
\label{na}
n_\mu \equiv -\(1/\sqrt{-g^{00}}\) \pa_\mu t ,
\ee
%
and $\kappa$ is an adjustable parameter that controls the damping
timescale.  Specifically, it is shown in~\cite{Gundlachetal} that
small constraint perturbations about Minkowski background decay
exponentially with a characteristic timescale of order $\kappa$.  We
note that the constraint damping term contains only first derivatives
of the metric and hence does not affect the principal (hyperbolic)
part of the equations.
\subsection{Regularization of the axis, $r=0$}
\label{sec_axis}
Having described the GH formulation, we now specialize to the
symmetric case.  We note first that in our coordinates (\ref{metric})
adapted to the symmetry the line element of the flat spacetime
becomes
\be
\label{flat1}
ds^2= - dt^2 + dr^2 +dz^2 +r^2 \,d\Omega_n^2, \ee
and that in this case the source function
(\ref{GH_coords}) does not vanish but becomes
\bea &&H^{\rm Mink}_\mu=-\Gamma^{\rm Mink}_\mu=\non &&=(0,
\frac{n}{r}, 0,(n-1)\,\cot \theta_1,\dots,\cot \theta_{n-1},0), \eea
where $\theta_i$ are angular coordinates of the sphere's line element
$d\Omega_n^2$.  Since near the axis a general spacetime is locally
flat, the radial component of the source function is generically
singular at $r=0$, diverging as $n/r$.  To regularize this radial
component, we thus subtract the singular background contribution by
transforming
\be
\label{H_reg}
H_\al \to H_\al +\delta^1_\al\,H_1^{\rm Mink} \ee
and prescribe gauge conditions using the regular sources.

Invariance of the line element (\ref{metric}) under the reflection
$r\rightarrow -r$ in our case implies that the metric components
$g_{01}$ and $g_{12}$ are odd functions of $r$, while $g_{00}, g_{11},
g_{22}, g_{02}, S$ and $\Phi$ are even in $r$.  The GH constraint
(\ref{GH_coords}) then dictates that $H_1$, regularized via (\ref{H_reg}) is
an odd function of $r$, while $H_0$ and $H_2$ are even in $r$.

Moreover, the requirement that the surface area of an $n$-sphere must
vanish at the axis\footnote{that is, that the radial and areal
  coordinates coincide at the axis, to avoid a conical singularity
  there.} implies $g_{11}(t,0,z)=\exp[2\,S(t,0,z)] $.  We note that
this is an extra condition on $S$, which thus has to satisfy both this
relation, as well as the constraint that it have vanishing radial
derivative at $r=0$---specifically that $g_{11}-\exp(2\,S) = O(r^2)$.
Therefore, at $r=0$ we essentially have three conditions on the two
fields $S$ and $g_{11}$.  In the continuum, and given regular initial
data, the evolution equations will preserve regularity, however, in a
finite-differencing numerical code this will be true only up to
discretization errors.  As a general rule-of-thumb, the number of
boundary conditions should be equal to the number of evolved variables
in order to avoid regularity problems and divergences of a numerical
implementation.

An elegant way to deal with this regularity issue involves definition
of a new variable, $\lambda$,~\footnote{We note that a similar
  variable was introduced in \cite{lambda_ref}, also for the purpose
  of regularization.}:
\be
\label{lambda_var}
\lambda \equiv \frac{g_{11} -e^{2\, S}}{r}.  \ee
At the axis one then has $\lam \sim O(r)$.  Therefore, after changing
variables from $S$ to $\lam$ by using $ S=(1/2) \log(g_{11}- r\,
\lam)$ in all equations, and imposing $\lam(t,0,z)=0$ at the axis, one
ends up with a system where there is no over-constraining due to the
demand of regularity at $r=0$. Crucially, we note that the
hyperbolicity of the GH system is not affected by the change of
variables.

While we note that a more straightforward regularization method that
maintains $S$ as a fundamental dynamical variable and employs
analytical Taylor-series expansion of the equations in the vicinity of
$r=0$ can also be used in simulations, its reliability degrades in the
strong field regime, where the regularization that uses $\lambda$ remains consistently
accurate.\footnote{ See, however,
  \cite{SorkinChoptuik,SorkinOren} for accurate simulations of the
  strong gravity regime in spherical symmetry where Taylor-series
  approach is used.}

\subsection{The field equations and Kaluza-Klein boundary conditions}
\label{sec_eqs}
With the metric ansatz (\ref{metric}) and the regularized source
function (\ref{H_reg}), our equations become 8 equations for 8
variables: six components of the 3-metric $g_{ab}$, and real and
complex scalars $\lam$ and $\Phi$ correspondingly. Schematically
the system can be written as
\bea -\half g^{cd} g_{ab,cd} &-& H_{(a,b)}+ \dots = \non &=& \pa_{(a}
\Phi \, \pa_{b)}
\Phi^*+ \frac{2}{D-2} g_{ab} V, \label{EqHgab} \\
-\half g^{cd} \lam_{,cd} &-&\half\,\frac{\pa_r\,H_1}{r}+\dots= \non
&=& \frac{2}{D-2}\, V\,\lam+\frac{|\pa_r \Phi|^2}{r}   , \label{EqHS} \\
g^{cd} \Phi_{,cd} &+& \dots = \pa V/\pa\Phi^*\label{EqHPhi}. \eea
Here ellipses denote terms that may contain the metric and it
derivatives and/or the source functions, in various combinations.
These equations are to be evolved forward in time starting from the
initial ($t=0$) time slice, where values for the fields and their
first time derivatives must be prescribed.

In order to completely specify the problem we have to provide boundary
conditions which the above equations are subject to. In this paper we
will be interested in a $D$-dimensional Kaluza-Klein spacetime of
topology ${\mathbb R}^{D-2,1}\times S_z^1$, where the $z$ direction is
considered periodic with asymptotic length $\h L$ namely that $z \sim
z +\h L$, and for the future use we also define half-period $L\equiv\h
L/2$. Asymptotically, this spacetime becomes Minkowski times the
compact circle and the corresponding boundary conditions are
\be
\label{asymp_bc}
g_{ab} \rightarrow \eta_{ab} ,~~~ \lam \rightarrow 0, ~~~\Phi
\rightarrow 0~~~H_a \rightarrow 0. \ee
%

\subsection{Coordinate conditions}
\label{sec_coord_condition}
As we have already mentioned, fixing the coordinates in the GH
approach amounts to specifying the source functions $H_a$. The choice
that we find to perform best in our case is a variant of the
damped-wave gauge (DWG) condition proposed recently in \cite{LS} (see
also \cite{PretoriusChoptuik_coll})
\be
\label{LS_Fa}
F_a=2\,\mu_1\,\log\(\frac{\gamma^p}{\al}\)\,n_a-2\,\mu_2\,\al^{-1}\,\gamma_{ai}\,\bt^{i},
\ee
where $\gamma_{ab}=g_{ab}+n_a n_b$ is the spatial metric whose
determinant $\gamma \equiv
det\,\gamma_{ij}=(g_{11}\,g_{22}-g_{12}^2)\,\exp(n\,S)$ has the factor
$r^n$ removed in accordance with the regularization (\ref{H_reg}), and
$\mu_{1,2}$ and $p$ are free parameters.\footnote{We note that the
  spatial part of DWG is essentially a version of the popular
  $\Gamma$-driver condition \cite{LS,gamma_dr}.} Since the gauge function (\ref{LS_Fa}) depends only on
the metric, we can simply set $H_a=F_a$ without destroying
hyperbolicity of our system.  Below we refer to this approach as the
algebraic DWG condition.

An alternative method that preserves the hyperbolicity was originally
devised by Pretorius for binary black hole simulations \cite{FP2,FP3}
and has also proven useful in the studies of the gravitational collapse of
scalar field in spherical symmetry \cite{SorkinChoptuik}.  This
strategy elevates the status of the $H_a$ to independent {\it
  dynamical} variables that satisfy time-dependent partial
differential equations.  The evolution equations for the $H_a$ are
designed so that the ADM kinematic variables---lapse $\al$ and
shift $\bt^i$ which (implicitly) result from the time
development---have certain desirable properties.  For example, the
equation for $H_0$ is tailored in an attempt to keep the value of the
lapse function of order unity everywhere---including near the surfaces
of the black holes---during the evolution.

One specific prescription for achieving this type of control evolves
the gauge source functions according to
\bea
\label{FP_gauge}
\Box H_t& =& -\xi_1 \,\frac{\al-\al_0}{\al^q} + \xi_2 \, H_{t,\mu}
n^\mu ,\non H_i &=&0, \eea
where $\Box$ is the covariant wave operator, and $\al_0, \xi_1, \xi_2$
and $q$ are adjustable constants.\footnote{In certain situations it is
  convenient to assume that $\xi_1$ and $\xi_2$ are given functions of
  space and time rather than mere constants.  For example, one might
  require that the gauge driver is switched on gradually in time, or
  that it be active only in certain regions, e.g.~in the vicinity of a
  black hole, and that its effect vanish asymptotically. } Thus the
temporal source function satisfies a wave equation similar to those
that govern the metric components in the system
(\ref{Eqs_constrdamp}).  The first term on the right-hand-side
of~(\ref{FP_gauge}) is designed to ``drive'' $H_t$ to a value that
results in a lapse that is approximately $\al_0$.  The second,
``frictional'' term tends to confine $H_t$ to this value.  For the
case of the spatial coordinates, Pretorius found that the simplest
choice of spatially harmonic gauge, $H_i=0,$ was sufficient in
simulations of binary black hole collisions.  A slight generalization
of this technique was considered in \cite{Scheel_etal} where instead
of using $H_i=0$, the spatial components of the source functions are
evolved according to
\be
\label{FP_gauge_m}
\Box H_i = -\xi_3 \,\frac{\bt_i}{\al^2} + \xi_2 \, H_{i,\mu}n^\mu \ee
where $\xi_3$ is an additional parameter.

Variants of gauge drivers were further investigated in the recent
\cite{Lindblom_etal_gauge,SorkinChoptuik,LS}. Specifically in
\cite{LS} the following hyperbolic first-order drivers were proposed
\bea
\label{LS_driver}
&&\pa_t H_a - \bt^i \pa_i H_a = -\nu\, (H_a-F_a) +W_a, \non &&\pa_t
W_a +\eta\, W_a = -\eta\, \bt^i \pa_i H_a, \eea
such that all time-independent solutions of this system satisfy
$H_a=F_a$, where $F_a$ are certain predetermined target gauge
functions, for instance (\ref{LS_Fa}), and the parameters $\nu$ and
$\eta$ are freely specified.

In Sec. \ref{sec_results_coords} we compare the performance of these
strategies and argue that in collapse situations the algebraic DWG approach
is the most robust of all. Namely, it does not require an extensive fine-tuning
of parameters, and it is long-term stable.
\subsection{Initial data}
\label{sec_id}
We now consider specification of initial data, which are values for
the fields and their first time derivatives at $t=0$.  For simplicity
we restrict attention to time-symmetric initial conditions for the
metric and $|\Phi|$.

Given this assumption, initial data for the scalar field reduces to
the specification of $\Phi(0,r,z)$, which we take to have the form of
a generalized Gaussian,
\be
\label{gaussian2d}
\Phi(0,r,z)=\Phi_0\,
e^{-\[(1-e_r^2)(r-r_0)^2+(1-e_z^2)(z-z_0)^2\]/\Delta^2} \ee
where $\Phi_0=|\Phi_0|\exp(i\,\varphi)$ is a complex amplitude and
$r_0,z_0,e_r,e_z,\Delta$ and $\varphi$ are real adjustable parameters,
supplemented by the choice
\be
\label{dtPhi_0}
\pa_t\Phi(0,r,z)=i\,\omega\,\Phi(0,r,z), \ee
that satisfies $\pa_t |\Phi|_{t=0}=0 $, where $\omega$ is another
parameter.

The momentum constraint is satisfied for our initial data, and writing
the initial metric as
\be
\label{conf_id}
ds^2=-\al^2 dt^2+\psi^4\(dr^2+dz^2+r^2 d\Omega^2_n\), \ee
one can show that the Hamiltonian constraint becomes an elliptic
equation for $\psi(0,r,z)$
\bea
\label{M0_id}
&&\pa_r^2 \psi +\pa_z^2 \psi + \frac{n}{r}\pa_r \psi +
\frac{n-1}{\psi}\[(\pa_r \psi)^2 +(\pa_z \psi)^2 \] +\non &&
+\frac{\psi}{2\, (n+1)} \Big[\half\(|\pa_r \Phi|^2 + |\pa_z \Phi|^2 +
\al^{-2} \psi^4 \omega^2 |\Phi|^2 \) + \non &&+ \psi^4 V \Big]=0, \eea
that is subject to the boundary conditions, $\pa_r \psi|_{r=0}=0,~
\pa_z \psi|_{z=0}=\pa_z \psi|_{z=L}=0$ and $\psi|_{r\to \infty}=1$.

We will assume initial harmonic coordinates, which implies, see (\ref{GH_coords}), that the
lapse can be determined in terms of $\psi$ as
\be
\label{al_psi}
\al(0,r,z)=\psi^{2\,n}.  \ee
After substituting this into equation (\ref{M0_id}) we solve for
$\psi$ and initialize our basic variables and their derivatives:
\bea
\label{id_vars}
g_{ij}(0,r,z)&=&\psi^4 \delta_{ij} ,~~ \lam(0,r,z)=0,\non \pa_t
g_{ij}(0,r,z)&=&\pa_t \lam(0,r,z)=0.  \eea
We finally note that in order to use the DWG conditions correctly, we
have to smooth out the transition from initially harmonic to later DWG
coordinates.  As described in Sec. \ref{sec_results_coords} this can
be done stably by multiplying $F_a$ in (\ref{LS_Fa}) by a
time-dependent factor that gradually grows from 0 to 1.  For the
driver version of DWG we also initialize $\pa_t W=W=0$ at $t=0$.
\subsection{Spacetime diagnostics}
\label{sec_diagnostics}
We employ several diagnostics in order to probe the geometry of the
spacetimes we construct.
\subsubsection{ Asymptotic charges: Mass and tension.}
\label{sec_asymptcharges}
Far away from an isolated system a natural radial coordinate is well
defined by comparison with the flat background, and the charges that
characterize the solution can be found from the asymptotic radial
behavior of the metric functions. In asymptotically flat spacetimes
with no angular momentum there is exactly one charge: the ADM mass.
However, in our system with the extra compact dimension, the tension,
associated with varying the length of the compact direction, can also
be defined.  A derivation of this result can be found in
\cite{KolSorkinPiran1,HO1}, but here we note why the appearance of an
additional charge could be expected.  Asymptotically the metric
becomes $z$-independent since from the lower-dimensional perspective the
$z$-dependent Kaluza-Klein modes are massive (with the discrete masses
$m_n=2\pi\,n/L,~ n>0$) and decay exponentially $\sim \exp(-m_n\,r)$.
In this situation asymptotic Kaluza-Klein reduction is possible with
the result that $g_{zz}$ behaves effectively as a scalar field that
carries an (unconserved) charge.  Designating the asymptotic fall-off
of the metric functions $g_{tt}$ and $g_{zz}$ as
\bea
\label{ab}
g_{tt}&=&-1+\frac{2\,a}{r^{D-4}},\non g_{zz}&=&1+\frac{2\,b}{r^{D-4}},
\eea the mass and tension of the solutions are defined as
\cite{KolSorkinPiran1,HO1}
\be \left[ \begin{array}{c} m \\ \tau\, \h L \\ \end{array} \right] =
{\Omega_{n} \over 8\pi(G_D/\h L)} \left[ \begin{array}{cc}
    n & -1 \\
    1   & -n \\
  \end{array} \right] \,
\left[ \begin{array}{c} a \\ b \\ \end{array} \right],
\label{asymp_to_charges}
\ee where $\Omega_n=2 \pi^{((n+1)/2)}/\Gamma[(n+1)/2]$ is the surface
area of a unit $n$-sphere, and ${\hat L}$ is the asymptotic length of the KK circle.
\subsubsection{ Apparent horizons}
\label{sec_ah_eh}
\begin{figure}[t!]
  \centering \noindent
  \includegraphics[width=5cm]{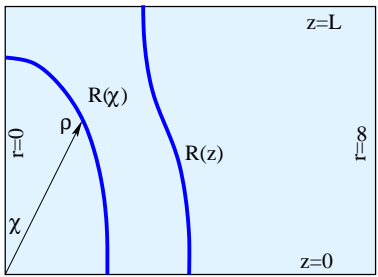}
  \caption[]{We are locating apparent horizons of two types: one,
  having a spherical topology, is conveniently parametrized by
    $\rho=R(\chi)$ (plotted here as centered at $(0,0)$), and another, having
    a cylindrical topology, is parametrized by $r=R(z)$. }
  \label{fig_ah_cartoon}
\end{figure}

It is well known that a process that concentrates sufficient
mass-energy within a small enough volume can lead to the formation of
a black hole.  In numerical calculations based on a space-plus-time
split, black hole formation is often inferred by the appearance of
apparent horizons.  An apparent horizon is defined as the outermost of
the marginally trapped surfaces, on which future-directed null
geodesics have zero divergence.  Specifically, in our case we will be
searching for simply connected horizons of two categories, see Fig.
\ref{fig_ah_cartoon}: (i) Quasi-spherical horizons of topology $S^n$,
in which case the surface describing the horizon can be taken as
\be
\label{ah_rhochi}
f(\rho,\chi)=\rho-R(\chi) \ee
where
\bea
\label{rho_chi}
\rho&=& \sqrt{(r-r_0)^2+(z-z_0)^2}\non
r-r_0&=& \rho \,\sin \chi \\
z-z_0&=& \rho \,\cos \chi \nonumber, \eea
where $(r_0,z_0)$ is the point where the horizon is centered, and (ii)
Horizons of topology $S^{n-1}\times S^1$ smeared along the $z$ direction, which do
not intersect the axis.  In this case a convenient parametrization is
\be
\label{ah_rz}
f(r,z)=r-R(z).  \ee
%
In either case we define an outward-pointing spacelike unit normal to
the surface $f=0$,
\be
\label{s_a}
s^\al= \frac{\gamma^{\al\bt}\,f_{,\bt}}{\sqrt{\gamma^{\rho\sigma}\,f_{,\rho}\,f_{,\sigma}} }.
\ee
The vanishing of the divergence, $\theta$, of the outgoing null rays
defined by $l^\al = s^\al +n^\al $ can be expressed as
\be \label{zero-divergence} \theta=(\gm^{\al\bt}-s^\al s^\bt)
\nabla_\al l_\bt = 0.  \ee

Substituting the expressions (\ref{ah_rhochi}) or (\ref{ah_rz}) into
(\ref{zero-divergence}) yields ordinary second order differential
equations for $R(\chi)$ or $R(z)$ correspondingly. The equations could
be solved by ``shooting'', subject to appropriate boundary conditions,
however, here we instead use the following point-wise relaxation
method, that is more suitable for parallel numerical implementations.
We start with supplying an initial guess for the entire function $R$ and
iterate the parabolic equation
\be
\label{ah_eqn}
\frac{\pa R}{\pa\tau}=-\theta(R'',R',R, g_{ab}, \pa g_{ab},x^a), \ee
in unphysical ``time'' $\tau$ until a solution is found. This equation
implies that depending on whether $R(\tau)$ is inside or outside the
horizon at any given moment, it will expand or shrink in the next
instant. In numerical simulations the hope is that the initial $R$
will ``flow'' to the apparent horizon in finite time. We find that in our case
this indeed happens when the initial guess is ``reasonably close'' to the final solution.

\section{Numerical approach}
\label{sec_numercal_approach}
Here we describe our strategy for the numerical solution of the GH
system (with a scalar matter source) in Kaluza-Klein spacetime.

\subsection{The numerical grid and the algorithm}
\label{sec_numerics}
We cover the $t$-$r$-$z$ space by a discrete lattice denoted by
$(t^n,r_i,z_j)=(n\,\Delta t,i\,\Delta r,j\,\Delta z)$, where $n,i$ and
$j$ are integers and $\Delta t$, $\Delta r$, and $\Delta z$ define the
grid spacings in the temporal and two spatial directions,
respectively. As described in the next section, the spatial domain is
compactified, and hence a grid of finite size $N_r \times N_z$ extends
from the axis to spatial infinity.  Approximations to the dynamical
fields, collectively denoted here by $Y$, are evaluated at each grid
point, yielding the discrete unknowns $Y^n_{i,j} \equiv
Y(t^n,r_i,z_j)=Y(n\,\Delta t,i\,\Delta r,j\,\Delta z)$.  In the
interior of the domain, the GH equations and the gauge-driver
equations are discretized using\footnote{Here $h$ stands for any of
  the mesh-spacings $\Delta t,\Delta r$ and $\Delta z$ that define our
  numerical grid.}  ${\cal O}(h ^2)$ finite difference approximations
(FDAs), which replace continuous derivatives with the discrete
counterparts, examples of which are given in Table \ref{table_FDA}. As
in \cite{FP1,FP2,SorkinChoptuik} our scheme directly integrates the
second-order-in-time equations.

\begin{table}[h!]
  \centering \noindent
  \hspace{-0.0in}
  \begin{tabular}{c||c}\hline
    & Centered derivatives \\ \hline \hline
    $\pa_t Y$& $(Y^{n+1}_{i,j}-Y^{n-1}_{i,j})/(2\Delta t)$    \\  
    $\pa_r Y$& $(Y^{n}_{i+1,j}-Y^{n}_{i-1,j})/(2\Delta r)$    \\  
    $\pa_{t}^2 Y $ &($Y^{n+1}_{i,j}-2\,Y^{n}_{i,j}+Y^{n-1}_{i,j})/(\Delta t)^2$   \\  
    $\pa_{r}^2 Y $ &($Y^{n}_{i+1,j}-2\,Y^{n}_{i,j}+Y^{n}_{i-1,j})/(\Delta r)^2$   \\  
    $\pa_{tr}^2 Y $ &($Y^{n+1}_{i+1,j}-Y^{n+1}_{i-1,j}-Y^{n-1}_{i+1,j}+Y^{n-1}_{i-1,j})/(4\,\Delta t \Delta r)$   \\ 
    $\pa_{rz}^2 Y $ &($Y^{n}_{i+1,j+1}-Y^{n}_{i-1,j+1}-Y^{n}_{i+1,j-1}+Y^{n}_{i-1,j-1})/(4\Delta r \Delta z)$ \\ 
    \hline
    & Backward derivatives \\ \hline \hline 
    $\pa_r Y$& $(4\,Y^{n}_{i+1,j}-3\,Y^{n}_{i,j}+Y^n_{i+2,j})/(2\Delta r)$    \\  
    $\pa_{r}^2 Y $ & $(2\,Y_{i,j}^n -5\, Y_{i+1,j}^n+4\,Y_{i+2,j}^n-Y_{i+3,j}^n)/(\Delta r)^2$ \\  
    \hline
  \end{tabular}
  \caption[]{Examples of second order finite-differencing approximation
    to derivatives calculated at a grid point $(n,i,j)$ that we use in our discretization scheme
    in the interior of the domain (centered stencil),
    and at the excision boundary (backward stencil) of the numerical grid.}
  \label{table_FDA}
\end{table}
Following discretization, we thus obtain finite difference equations
at every mesh point for each dynamical variable.  Denoting any single
such equation as
\be
\label{FDA_eq}
{\cal L}_Y|^n_{i,j}=0.  \ee
we then iteratively solve the entire system of algebraic equations as
follows.

First, we note that for those variables that are governed by equations
of motion that are second order in time, our ${\cal O}(h^2)$
discretization of the equations of motion results in a three level
scheme which couples advanced-time unknowns at $t^{n+1}$ to known
values at retarded times $t^n$ and $t^{n-1}$.  In order to determine
the advanced-time values for such variables, we employ a point-wise
Newton-Gauss-Seidel scheme: starting with a guess for $Y^{n+1}_{i,j}$
(typically, we take $Y^{n+1}_{i,j}=Y^n_{i,j}$) we update the unknown
using
\be
\label{Gauss-Seidel} {Y^{n+1}_{i,j}} \rightarrow {Y^{n+1}_{i,j}} -
{\R_Y|^n_{i,j} \over\J_Y|^n_{i,j}}. \ee
Here, $\R_Y $ is the residual of the finite-difference
equation~(\ref{FDA_eq}), evaluated using the current approximation to
$Y^{n+1}_{i,j}$, and the diagonal Jacobian element is defined by
\be
\label{jacobian}
\J_Y|^n_{i,j} \equiv \frac{\pa {\cal L}_Y|^n_{i,j}}{\pa Y^{n+1}_{i,j}}.  \ee
In the cases where we used gauge drivers (\ref{LS_driver}) we found
that an iteration based on the Crank-Nicholson discretization scheme of
the corresponding first order equations performed well. Specifically,
writing any such equation schematically as $\dot Y = f_Y(Y,\pa Y
,\dots)$, we update using
\be
\label{CN}
Y^{n+1}_{i,j} \rightarrow Y^{n}_{i,j} +\half \Delta t\,
\(f_Y|^{n+1}_{i,j}+ f_Y|^{n}_{i,j}\).  \ee
We iterate (\ref{Gauss-Seidel}) and (\ref{CN}) over all equations
until the total residual norm, see (\ref{res}), falls below a desired
threshold.

In order to inhibit high-frequency\footnote{``High-frequency'' refers
  to modes having a wavelength of order of the mesh spacings, $\Delta
  r$ and $\Delta z $.}  instabilities which often plague FDA
equations, our scheme incorporates explicit numerical dissipation of
the Kreiss-Oliger (KO) type \cite{KO}.  Following \cite{FP1}, at the
interior grid points, $\{(i,j) |~ 2\leq i \leq N_r-2,~ 2\leq j \leq
N_z-2 \}$, and for each dynamical variable we apply a low-pass KO
filter by making the replacement
\bea
\label{KO_filter}
Y_{i,j} \to&& Y_{i,j}-\eps_{\rm KO}\,d_{i,j} \non d_{i,j} \equiv&&
\frac{1}{16}\Big(Y_{i-2,j}-4\,Y_{i-1,j} -4\,Y_{i+1,j} + Y_{i+2,j}+
\non &&~~+Y_{i,j-2}-4\,Y_{i,j-1} -4\,Y_{i,j+1} + Y_{i,j+2}+\non
&&~~+12\,Y_{i,j} \Big)\eea
at both the $t^{n-1}$ and $t^n$ time-levels before updating the
$t^{n+1}$ unknowns. Here $\eps_{\rm KO}$ is a positive parameter
satisfying $0\le \eps_{\rm KO} \le 1$ that controls the amount of
dissipation.  An extension of the dissipation to the boundaries
\cite{FP1} was also tried, but this has not resulted in any
significant improvement of the performance of the code.

\subsection{Coordinates and boundary conditions}
\label{sec_coords_bc}

While the physical, asymptotically flat (times a circle, in our case)
spacetime extends to spatial infinity, in a numerical code one can
only use grids of finite size.  Instead of using a standard strategy
that deals with this issue by introducing an outer boundary at some
finite radius where approximate boundary conditions are imposed, we
adopt another technique---proven successful in previous work in
numerical relativity, see
e.g.~\cite{FP2,SorkinChoptuik,Choptuik_BS}---and compactify the
spatial domain. We find that compactifying the radial direction and
imposing the (exact) Dirichlet conditions (\ref{asymp_bc}) at the edge
of the domain works well, provided that we use sufficient dissipation.
In particular, it is known that due to the loss of resolution near the
compactified outer boundary (assuming a fixed mesh spacing in the compactified coordinates), 
outgoing waves generated by the dynamics
in the interior will be partially reflected as they propagate toward
the edge of the computational domain, and these reflections will then
tend to corrupt the interior solution.  By adding sufficient
dissipation one can damp the waves in the outer
region, attenuating any unphysical influx of
radiation. This enables one to use the compactification meaningfully.

The results presented in this paper were obtained using the
compactification of the form
\be
\label{comact_r}
\tilde{r}=\frac{r}{1+r}, \ee
where the compactified $\tilde{r}$ ranges from $0$ to $1$ for values
of the original radial coordinate $r \in [0,\infty)$. In practice, we
define a uniform grid in the compactified $\tilde r$ and use
chain-rule to replace derivatives with respect to $r$ with the
derivatives with respect to $\tilde r$ in all dynamical equations.
The asymptotic boundary conditions (\ref{asymp_bc}) at $\tilde{r}=1$
are then imposed exactly: $g_{ab} =\eta^{(3)}_{ab}, \lam= 0, \Phi= 0$,
and $H_a=W_a =0$.

We have previously described the boundary (regularity) conditions at
$\tilde{r}=r=0$ in Sec.~\ref{sec_axis}.  Denoting by $Y^{n+1}_{1,j}$,
for $j=2,\dots,N_z-1$, the advanced-time value at the axis for any of
the variables, $g_{00}, g_{11}, g_{22}, g_{02}$,$H_0$, and $H_2$ that
have vanishing derivative at $r=0$, we use the update
$Y^{n+1}_{1,j}=(4\,Y^{n+1}_{2,j}-Y^{n+1}_{3,j})/3$, which is based on
a second-order backwards difference approximation (see
Table ~\ref{table_FDA}) of $\partial_r Y=\partial_{\tilde r}Y=0$.  For
the quantities $g_{01}, g_{12}, \lam$ and $H_1$, which are odd in $r$
as $r\to0$, we simply use $Y^{n+1}_{1,j}=0$.

For simplicity, in this paper we consider only configurations that
have reflection symmetry about $z=0$ which together with periodicity
in z, implies
\be
\label{z_bc}
\pa_z Y|_{z=0}=\pa_z Y|_{z= L}=0.  \ee
We update points at $z=0$ and $z=L$ using backward difference
approximation similar to that in Tab.~\ref{table_FDA}. In particular
for $i=2,\dots,N_r-1$, we use
$Y^{n+1}_{i,1}=(4\,Y^{n+1}_{i,2}-Y^{n+1}_{i,3})/3$ at $z=0$, and
$Y^{n+1}_{i,N_z}=(4\,Y^{n+1}_{i,Nz-1}-Y^{n+1}_{i,Nz-2})/3$ at $z=L$.

\subsection{The elliptic equation of initial data}
\label{sec_eqs_id}
It turns out that generally the equation (\ref{M0_id}) for the
conformal factor is ill-posed, namely that the linear equation
governing small perturbations about a solution of (\ref{M0_id}) does
not admit a unique solution for given boundary conditions \cite{York}.
Therefore, an attempt to solve (\ref{M0_id}) using standard relaxation
methods will generically fail, as the relaxation is not guaranteed to
converge.  However, a method circumventing this difficulty exists
\cite{York}, and this is through a rescaling $\Phi=\hat\Phi \, \psi^s$
that transforms the Hamiltonian constraint (\ref{M0_id}) into
\bea
\label{M0_id_resc}
&&\pa_r^2 \psi +\pa_z^2 \psi + \frac{n}{r}\pa_r \psi +
\frac{n-1}{\psi}\[(\pa_r \psi)^2 +(\pa_z \psi)^2 \]+ \non
&&\frac{s^2\psi^{2\,(s-1/2)}}{4(n+1)}|\hat \Phi|^2
\[(\pa_r \psi)^2 +(\pa_z \psi)^2 \] +\non
&&\frac{s \psi^{2\,s}}{4(n+1)}\(\pa_r |\hat \Phi|^2 \pa_r \psi +\pa_z |\hat \Phi|^2 \pa_z \psi\)+\\
&& \frac{\psi^{2\,(s+1/2)}}{4\, (n+1)} \(|\pa_r \hat \Phi|^2 + |\pa_z
\hat \Phi|^2\) +\non && \frac{\omega^2 \psi^{2\,(s-2\,n+5/2)}}{4\,
  (n+1) } |\hat \Phi|^2+ \frac{\psi^{5}}{2\,(n+1)}\, V(
|\Phi|\,\psi^{s} ) =0.\nonumber \eea
By choosing the power $s$ such that terms in this equation which are
proportional to $\psi^{p_i}$ have only non positive $p_i$'s one
renders the problem well posed \cite{York}.  In the case of a free
scalar field having the potential $V \propto |\Phi|^2$, which we
assume here, this implies that $s<-5/2$, see a related discussion in
\cite{SorkinPiran}.

Note that the physical data are $\Phi$, not $\h \Phi$, and therefore we
solve equation (\ref{M0_id_resc}) in an iterative manner where we
start with an initial guess $\psi=\psi_0(r,z)$ and at each iteration
$i>1$ update $\h \Phi_{i+1} = \Phi\,\psi_{i}^{-s}$ that is then used
in (\ref{M0_id_resc}) in order to solve for $\psi_{i+1}$.  Most of the
results obtained in this paper were generated using $\psi_0=1$ and
$s=-3$.  Interestingly, it turns out that provided the initial
guess is not too distant from the solution, the original equation
(\ref{M0_id}) is numerically stable without the rescaling. This happens, for
instance, in the weak field regime where the guess $\Psi_0=1$ relaxes
to the solution without any trouble.

\subsection{Excision}
\label{sec_excision}
We use excision to dynamically exclude from the computational domain a
region interior to the apparent horizon that would eventually contain
the black hole singularity.  This approach relies on the observation
that in spacetimes that satisfy the null energy condition and assuming
the cosmic censorship holds, the apparent horizon is contained within
the event horizon.  This ensures that the excluded region is causally
disconnected from the rest of the domain (see
\cite{thornburg_excision} and the references therein for further
discussion).  Operationally, once an apparent horizon is found, we
introduce an excision surface, $R_{\rm EX}$, contained within the apparent horizon,
 and such that all characteristics at $R=R_{\rm EX}$
are pointing inwards. This specific
characteristic structure eliminates the need for boundary conditions
at $R_{\rm EX}$: rather, advanced-time unknowns located on the
excision surface are computed using finite difference approximations
to the interior evolution equations, but where centered difference
formulas are replaced with the appropriate one-sided expressions given
in Table \ref{table_FDA}.

Currently our apparent horizon finder is only capable of locating horizons of the shapes
depicted in Fig. \ref{fig_ah_cartoon}. In this case the radius of the excision surface typically satisfies $R_{\rm
  EX}\ltsim 0.7 R_{\rm AH}$, where $R$ is the coordinate radius.
\section{Performance of the code}
\label{sec_num_experements}
\begin{figure}[t!]
  \centering \noindent
  \includegraphics[width=7cm]{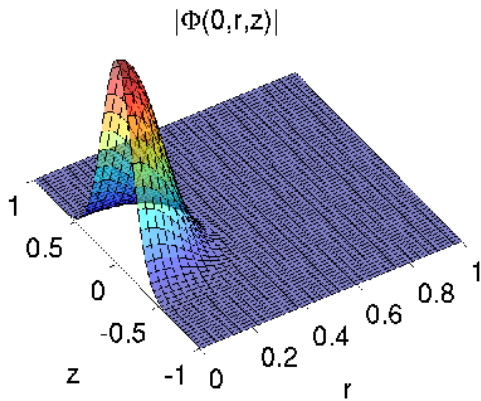}
  \caption[]{We choose the initial distribution of the scalar field as
    a generalized Gaussian (\ref{gaussian2d}) and assume that the initial time derivative
    is given by (\ref{dtPhi_0}). We mostly consider localized initial
    pulses (such as the one shown here), that are obtained by setting
    $e_r=e_z=0$ in (\ref{gaussian2d}).  The radial
    direction is compactified, and the hypersurfaces $z=1$ and $z=-1$
    identified in a Kaluza-Klein spacetime under consideration.}
  \label{fig_Phi_t0}
\end{figure}

In this section we investigate the performance of the code in series
of simulations that evolve regular initial distribution of complex
scalar field in five and six dimensional Kaluza-Klein spacetime.
The code uses pamr/amrd infrastructure
\cite{pamr_amrd} where our suitably interfaced numerical routines are
called by the amrd driver.  All our results are generated using an
initial scalar field profile of the Gaussian form~(\ref{gaussian2d}),
with fixed values $r_0=z_0=0$ and $\Delta=0.25$, so that the scalar
pulse is always initially centered at the axis, see Fig.
\ref{fig_Phi_t0}.
In addition, we set $e_r=0$ and use $e_z=0$, unless otherwise specified. We choose the
initial frequency in (\ref{dtPhi_0}) to be $\omega=20$, and the asymptotic size of
the KK circle, ${\hat L}=2$. The scalar
field potential used here is taken to be of the form
$V(\Phi)=m_\Phi^2\,|\Phi|^2$, where without
much loss of generality we set $m_\Phi=1$.

Because we mostly use a time-explicit finite difference scheme, we
expect restrictions on the ratio $\lambda_C \equiv \Delta t/h$ (the
Courant factor) that can be used while maintaining numerical
stability.  In the results discussed below we chose $0.2\lesssim
\lambda_C \lesssim 0.5$ in the weak-field regime, and
$0.05\lesssim\lambda_C\lesssim 0.25$ for evolving the strong data.
Taking larger $\lam_C$ usually leads to amplification and dominance of
numerical errors near the axis, that shows up as diverging high frequency
oscillations. In most cases we use uniform grids of the
same size in both spatial directions, $h \equiv \Delta r=\Delta z$, and
our lowest and highest resolution simulations have $h = 1/16$ and $h = 1/256$, respectively.
The highest-resolution runs generally required $\lambda_C \leq 0.2$ for
stability.

As expected, the outcome of the collapse depends on the strength of
the initial data. Low density distributions describe weakly
gravitating scalar pulses which completely disperse in all instances.
Strong data generate spacetimes in which black holes form, or
almost form.  For fixed $\Delta, e_r, e_z, m_\phi, {\hat L}$ and $\omega$
a single free parameter that controls the strength of the pulse---and
hence, the outcome of the evolution---is the initial amplitude $\Phi_0$.
In this case ``strong initial data'' will refer to situations when an increase
of the initial amplitude by less than $10\%$
leads to formation of a curvature singularity, and the other data will be called ``weak''.
The critical amplitude is at the threshold for the singularity formation.

For each spacetime that we construct, the asymptotic mass and tension are
computed using (\ref{asymp_to_charges}), where the constants $a$ and $b$ are found by fitting the 
metric components $g_{tt}$ and $g_{zz}$ with the functions of the form $g(r)=g_\infty+g_1/r^n+g_2/r^{n+1}$ 
in the asymptotic region. The errors in the constants are determined by the fitting 
uncertainty which in our case is $\sim 1-2\%$ (larger for weaker data). We find that our 
initial data sets usually have $a \simeq n\,b$, and it follows from (\ref{asymp_to_charges}) that $\tau \simeq 0$, 
within the numerical accuracy of our method.           

\begin{figure}[t!]
  \centering \noindent
  \includegraphics[width=8.8cm]{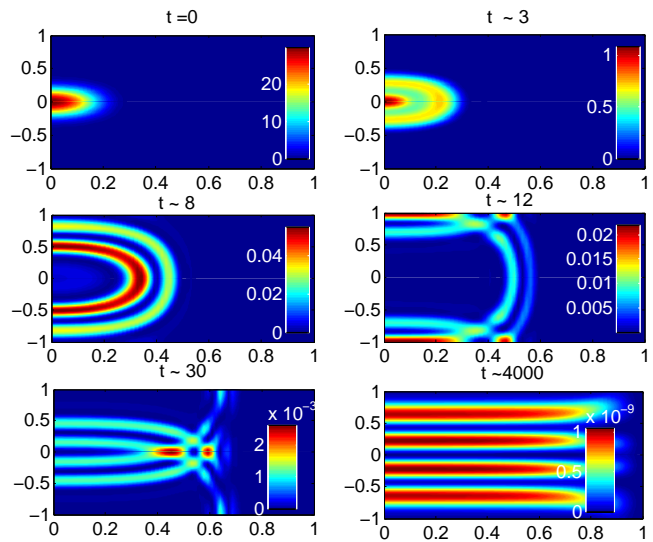}
  \caption[]{Snapshots of the evolution of energy-momentum density
    $\rho=n^a\,n^b\, T_{ab}$ in the simulation of the initial data
    defined by $\Phi_0= 0.35\,(1-i) $ in 5D, using algebraic DWG.  The
    horizontal axis represents the compactified radial direction, and the vertical
    axis is along the periodic $z$-direction. The lapse of time is measured in units of the total mass.
    The initial collapse of the pulse around the center $(r,z)=(0,0)$ is accompanied by a growth of
    the amplitude of $\rho$  by about an order of magnitude.  The
    pulse then bounces off and several outgoing shells form.  The shells propagate along the periodic
    $z$-direction and interact, creating a typical interference pattern.  As expected, the $z$-dependence vanishes in the
    asymptotic regions.  In the later stages of the evolution $\rho$ is small, and the frequency spectrum of the
    oscillations along $z$ is dominated by a few lowest eigenfrequencies
    associated with the compact KK circle.}
  \label{fig_rho_t_035}
\end{figure}
A useful quantity that illustrates energy-momentum distribution is the
density $\rho =T_{ab} n^a n^b$.  Figure \ref{fig_rho_t_035}
depicts $\rho$ at several moments during the 5D evolution of weak initial
data defined by $\Phi_0= 0.35\,(1-i) $, using algebraic damped wave gauge (DWG), see (\ref{LS_Fa}).
Figure \ref{fig_rho_t_035} shows that in the early stages of the evolution
the pulse (initially localized around the center
$(r,z)=(0,0)$) undergoes a gravitational collapse that is accompanied by a growth of the central
energy-density. At a later time, however, the pulse bounces off while forming several shells, and disperses.
The distribution of the energy density is
anisotropic inside the shells, with a higher concentration of
energy occurring along the axis, $r=0$, and at the equator, $z=0$. The
matter waves that travel along the compact KK circle collide to form a typical
interference picture (see Fig. \ref{fig_rho_t_035}) and the pattern becomes
increasingly complicated in the course of time, when more and more waves undergo interaction.
As expected in the KK background the $z$-dependence of solutions vanishes in the asymptotic region, since
the $z$-dependent modes are massive and fall off exponentially fast.

\begin{figure}[t!]
  \centering \noindent
  \includegraphics[width=7.0cm]{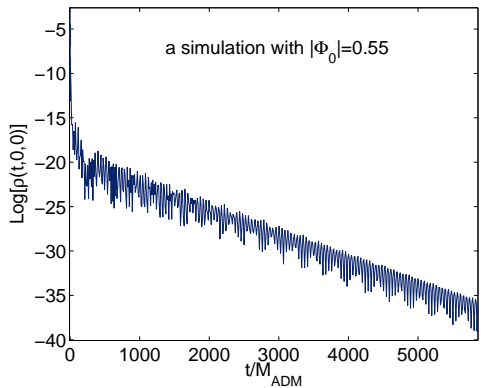}
  \caption[]{The logarithm of the central energy-density, $\rho$, in a five-dimensional weak
    field simulation defined by $|\Phi_0|=0.55$ that uses algebraic
    DWG. The resulting spacetime has the mass $M = 0.67 \pm 0.03$ and the tension
    is zero within the numerical accuracy. In the early stages of the evolution, the pulse collapses and $\rho$
    grows. Subsequently, the energy density decreases following a power-law dependence
    during the period lasting from $t\sim{\cal O}(10) M$ until $t\sim{\cal O}(100) M$.
    After that $\rho$ decays exponentially
    $~\exp[-t/( 800 M)]$.  The high-frequency
    oscillations are associated with
    the eigenmodes of the compact KK circle.  }
  \label{fig_rho_r0}
\end{figure}
In order to obtain more quantitative insight into the process, we plot
in Fig. \ref{fig_rho_r0} the evolution of the logarithm of $\rho$ at
the center.  During the highly dynamical early epoch,
lasting until $\sim{\cal O}(10) M$, the field collapses, bounces
off the center, spreads along the $z$-direction, and starts
dispersing to infinity.  In the first stage of the dispersion,
lasting until $\sim{\cal O}(100) M$, the decay of $\rho$ follows a
power-law, and in the late times the decay is exponential with a
characteristic time scale of a few hundreds of $M$. The maximal scalar
curvature is attained during the initial collapse
phase; in the shown simulation the curvature reaches the
magnitude of order of $\sim{\cal O}(100)$ (in units of $M^{-2}$).  The discrete spectrum of
the high-frequency oscillations consists of the normal frequencies
$f_n \sim n/L,~ n =1,2,\dots$ defined by the size of the KK circle.
We observe that the higher frequency modes decay faster, and the late-time 
spectrum is dominated by a few lowest frequency modes, see also bottom
right panel of Fig. \ref{fig_rho_t_035}.

\begin{figure}[t!]
  \centering \noindent
  \includegraphics[width=7.5cm,height=6.0cm]{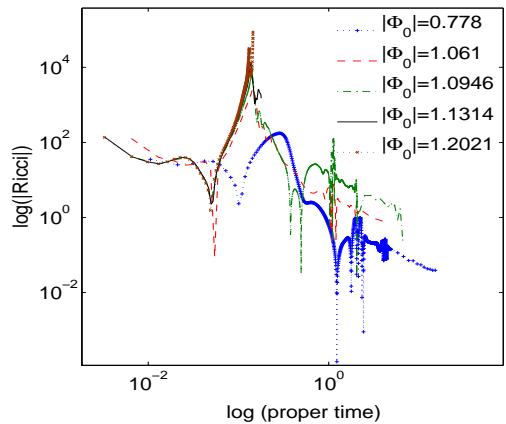}
  \caption[]{A log-log plot of the Ricci scalar (in units of $M^{-2}$) as a function of the
    proper time, $t_{\rm prop} \equiv \int_0^t \al\, dt$, both evaluated at the center, in several
    simulations in $5D$. The first peak in the curves corresponds to the
    initial collapse phase, and for stronger data the Ricci scalar
    diverges in finite time, signaling a curvature singularity (case
    $|\Phi_0|=1.2021$).  In subcritical cases, the curvature decreases
    after the initial growth, and the secondary peaks
    correspond to collisions of the shells traveling along the KK circle
    and arriving back at $(0,0)$.}
  \label{fig_maxRtprop}
\end{figure}
When the initial amplitude of the scalar field increases, the
maximal curvature achieved during the collapse grows, and when $|\Phi_0|$ surpasses a certain
threshold, the curvature diverges, signaling the appearance of a singularity; see Fig.
\ref{fig_maxRtprop}. Currently we are able to estimate the critical
amplitude with a relatively low precession of approximately 1 part in 800.
In five dimensions the amplitude is between $1.0946$ and $1.096$ such that initial data with
$|\Phi_0|\leq 1.0946$ completely disperses, and the data $|\Phi_0|\geq 1.096$ gives rise to
curvature singularities and black holes.

\begin{figure}[t!]
  \centering \noindent
  \includegraphics[width=8.8cm,height=9.5cm]{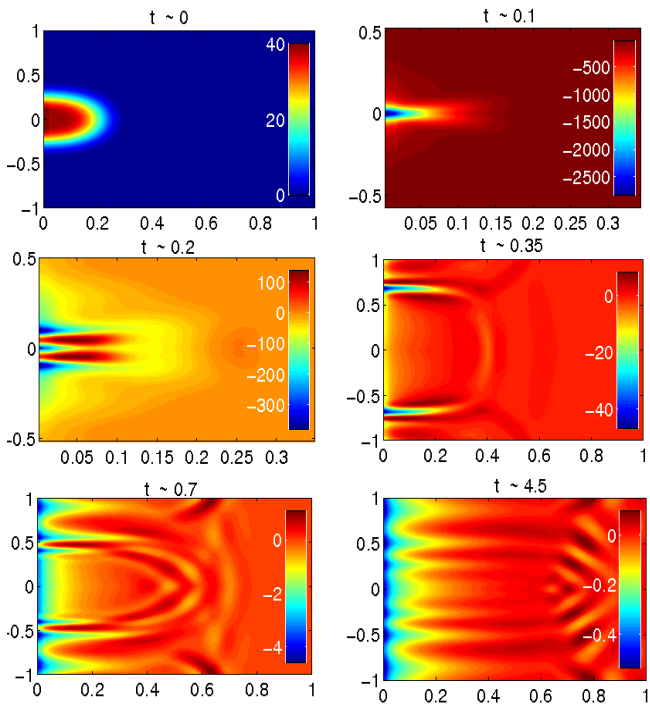}
  \caption[]{ The Ricci scalar at several instants of the evolution of the initial data
    characterized by $\Phi_0=0.774(1-i)$, currently our
    strongest subcritical data set. The numerical resolution used in this
    simulation is $129\times 129$.
    After the initial collapse stage an expanding shell of matter forms. The energy-density distribution
    inside the shell quickly becomes anisotropic with most of the energy localized within individual
pulses traveling along the axis and inside the pulse emitted radially along the equator, $z=0$.
The pulses at the axis undergo interactions and spread the energy-density along $r=0$.
In late stages of the evolution, all the matter is radiated away to infinity.  }
  \label{fig_Ric_t_0774}
\end{figure}

A covariant way to illustrate the distribution of matter
and the geometry of the spacetime is provided by the Ricci\footnote{In this and other figures Ricci scalar is 
measured in units of $M^{-2}$.} scalar, and
Fig. \ref{fig_Ric_t_0774} shows its evolution computed in the
evolution of our strongest subcritical data set,
$|\Phi_0|=|0.774(1-i)|\simeq 1.0946$. The initial pulse collapses,
bounces off and forms several shells, that start dispersing after
$t\sim 0.2 M$.  While the anisotropy of the energy distribution inside
the shells is small for weak data, it is obvious in the strong field regime.
Some matter is ejected along the equator, $z=0$, and two
distinctive pulses, moving in the opposite directions, form at the axis.
The pulses collide and interact, which results in smearing of the energy-density along the axis,
see bottom right panel of Fig. \ref{fig_Ric_t_0774}. The distribution is not
stationary, rather the matter is continuously leaking to infinity,
and the late-time behavior is qualitatively similar to the weak field case,
shown in Fig. \ref{fig_rho_t_035}.
\begin{figure}[h!]
  \centering \noindent
 \includegraphics[width=8.8cm,height=9.8cm]{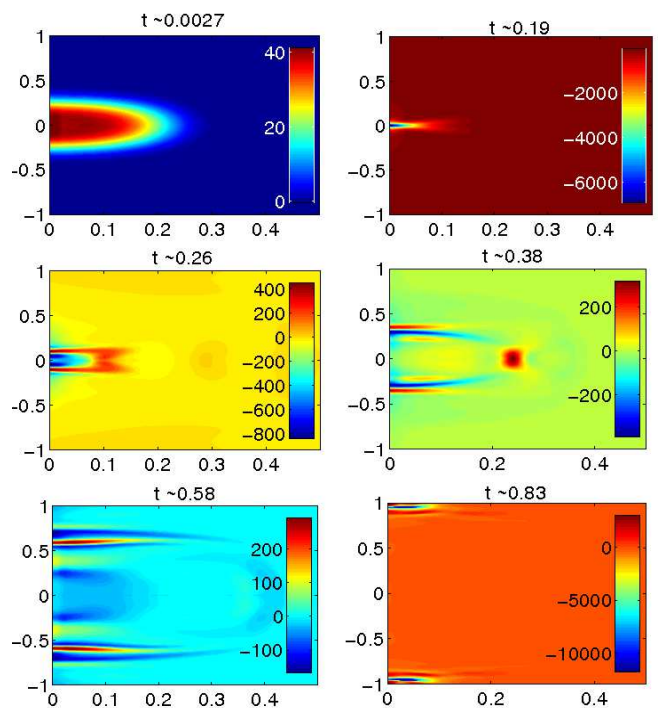}
  \caption[]{Snapshots of the Ricci scalar at several instants of the
    evolution of the initial data characterized by
    $\Phi_0=0.775(1-i)$ at the resolution of $257\times 257$.
    The matter distributes anisoropically inside the bouncing shells
    such that two localized  pulses, traveling along the axis, form. When the pulses
    collide at $z=1$ a curvature singularity appears.}
  \label{fig_Ric_t_0775}
\end{figure}
\begin{figure}[t!]
  \centering \noindent
  \includegraphics[width=8.8cm,height=9.8cm]{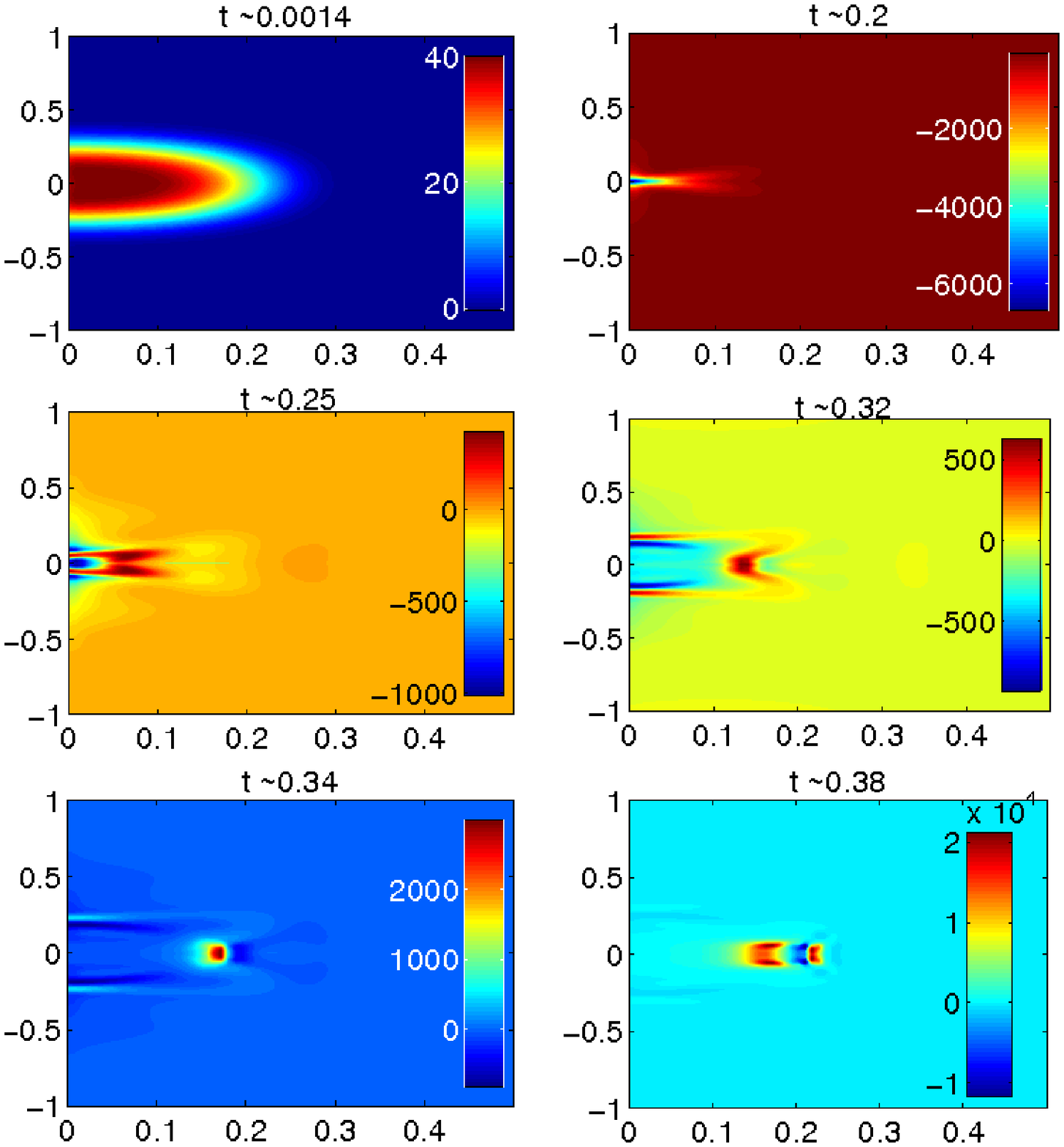}
  \caption[]{The Ricci scalar at several instants during the
    evolution defined by $\Phi_0=0.777(1-i)$. A curvature singularity forms along the equator,
    inside the lump of matter emitted in the radial direction during the initial collapse-bounce process.
    This simulation uses the resolution of $257\times 257$.  }
  \label{fig_Ric_t_0777}
\end{figure}

It turns out that for higher initial amplitudes, $|\Phi_0| \gtsim 1.096$,
the pulses that form at the axis are able to individually collapse and develop
curvature singularities, indicated by diverging scalar curvature and
energy-momentum density while the lapse remains finite (of order one)
everywhere. Presently, our horizon-finder is not designed to locate
the moving apparent horizons that may arise in this case around the singularities,
and it would be very interesting to verify whether or not such engulfing horizons indeed form.
The time when the curvature singularities appear depends
on the strength of the initial data in the manner that the closer to threshold we
are the later into the evolution the curvature diverges. For instance, in
the case of $|\Phi_0|=1.096$ the pulses, collapse and become singular
as they cross the circle and collide at $z \simeq 1$, see Fig \ref{fig_Ric_t_0775}.
However, for $|\Phi_0|=1.0975$, the curvature inside each pulse blows up earlier, when
they reach $z\sim 1/2$.

Intriguingly, we observe that for even stronger initial data formation
of the curvature singularity ensues differently.  Specifically, the pulse of matter
which is usually emitted during the initial collapse-bounce stage outwards
along the equator, is now seen to also be able to collapse and develop a singularity.
Snapshots of the process are shown in Fig. \ref{fig_Ric_t_0777} that was obtained
in the evolution of the data defined by $|\Phi_0|=1.099$.  Since all our attempts
to detect an apparent horizon, engulfing the singular region and the center,
have failed we believe that the horizon, if it forms in this case,
must be localized around the moving curvature singularity.

\begin{figure}[t!]
  \centering \noindent
  \includegraphics[width=7cm]{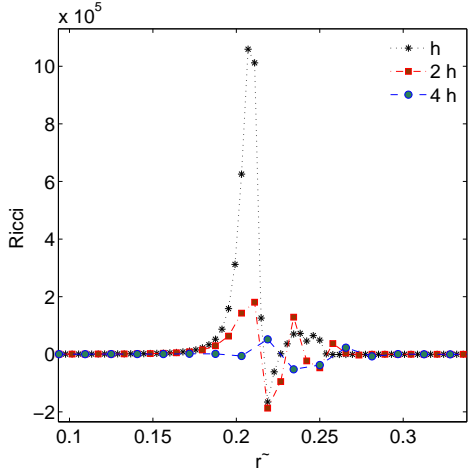}
  \caption[]{The Ricci scalar (in units of $M^{-2}$) along the equator at $t\sim 0.37 M$, shortly before 
    the evolution defined by $\Phi_0=0.777(1-i)$ develops a singularity at ${\tilde r} \sim 0.21$. 
    The maximal value that the Ricci scalar
    attains in this evolution depends on the resolution (here $h=1/256$), being larger on finer numerical meshes, which 
   indicates the geometrical nature of the emerging singularity. }
  \label{fig_divergenceR}
\end{figure}

In Fig. \ref{fig_divergenceR} we plot the Ricci scalar along the equator at $t\sim 0.37 M$, shortly before 
the appearance of a singularity at ${\tilde r} \sim 0.21$ causes the code to break down. Figure \ref{fig_divergenceR}
shows that the maximal value of the Ricci scalar along this slice is determined by the numerical 
resolution: the finer meshes we use, the larger the values attained. This signals emergence of a curvature 
rather than a coordinate singularity.  
\begin{figure}[t!]
  \centering \noindent
  \includegraphics[width=8.8cm,height=9.5cm]{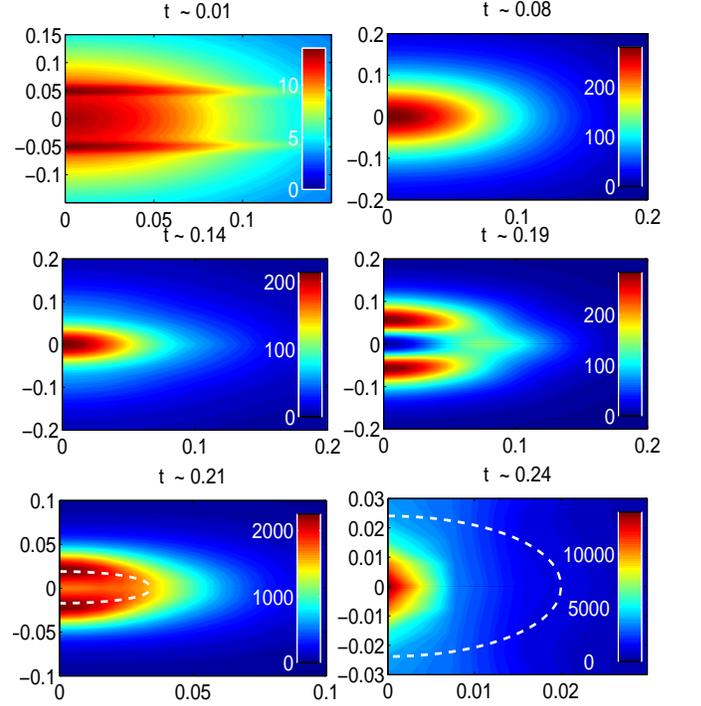}
  \caption[]{The distribution of energy-density in the evolution of the
    initial data, defined by $\Phi_0= 0.8\,(1-i) $ in a simulation that uses the resolution of $201\times 201$ .
    A strong gravitational interaction in this case precludes any significant amount of matter from escaping to infinity and leads to
    formation of a black hole soon after the first bounce. The black hole is centered
    around $(0,0)$ and is signaled by the appearance of an apparent
    horizon (designated by a thick dashed line) and an explosive growth of the scalar curvature and the energy-density.}
  \label{fig_rho_t_08}
\end{figure}

Figure \ref{fig_rho_t_08} shows snapshots of the energy-density
distribution obtained in a supercritical simulation initiated by $\Phi_0=0.8(1-i)$.
The resulting spacetime has the total mass $M=2.40\pm 0.06$ and the tension $\tau
\simeq 0$.  The initial stages of this evolution are qualitatively
similar to those of weaker data, however, after the first bounce
off the center (at $t \simeq 0.19 M$) the pulse recollapses and a
black hole forms as indicated by the appearance of an apparent horizon.
The energy-density $\rho$ and scalar curvature both blow up in the vicinity of
$(r,z)=(0,0)$ in finite time, while the lapse and shift remain finite.

In order to locate apparent horizons we solve (\ref{ah_eqn}) iteratively starting
with some initial guess for the entire function that parametrizes the horizon.
In the case of $\Phi=0.8(1-i)$ we are searching for a
horizon that has spherical topology. We use the parametrization (\ref{ah_rhochi}) with $r_0=z_0=0$ and solve (\ref{ah_eqn})
initialized by $R=0.1$. Setting the target accuracy to
$\sim 10^{-3}$ and using $100$ grid points to represent the horizon,
we are able to solve the equation in $\sim 1000$ iterations. The resulting horizon
is shown in bottom panels in Fig. \ref{fig_rho_t_08}.

After an apparent horizon is found we use excision to remove a region inside the
horizon that contains curvature singularity. However, it turns out that regardless of our
specific gauge choice the lapse keeps evolving inside the horizon, and continuously decreases
until it is reaching the magnitudes of order $\sim 10^{-5}$ near
the excision boundary. In this situation truncation errors in quantities near $R_{EX}$
occasionally cause the computation of non-positive values for the
lapse, which immediately leads to code failure. Unfortunately, this happens when the
apparent horizon is still fairly dynamical and continues to change its shape and size.
Therefore, we are currently not able to determine the stationary black hole state in
the end of the evolution.

\begin{figure}[t!]
  \centering \noindent
  \includegraphics[width=8.7cm,height=9.5cm]{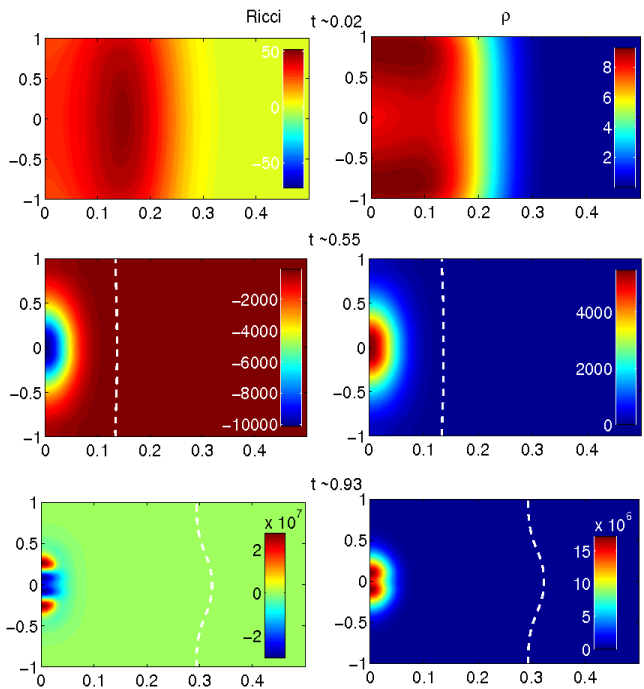}
  \caption[]{Distributions of the energy-density (on the right) and
  the Ricci scalar (on the left) computed in the evolution of a nearly uniform along $z$
  initial data defined by $\Phi_0= 1.2\,(1-i)$, and $e_r=0,e_z=0.995$, that uses the resolution of $101\times 33$.
     A strong gravitational interaction leads to a rapid formation of
     a black string, signaled by the appearance of an apparent horizon of cylindrical topology
     (shown as a thick dashed line). Note, that the matter and the curvature inside the horizon are strongly
     localized around the center $r=z=0$. }
  \label{fig_rhoRic_t_12}
\end{figure}
Having described the low mass configurations we will now briefly discuss more massive solutions of certain type.
One initial configuration that we have evolved consisted of a nearly uniform along $z$ distribution
of the matter---achieved by setting $e_z=0.995$ in (\ref{gaussian2d})---with the initial amplitude of $\Phi_0=1.2(1-i)$.
The Ricci scalar and the energy-density computed in this evolution are shown in Fig. \ref{fig_rhoRic_t_12}
together with the resulting apparent horizon that appears at $t\sim 0.55$ and has a cylindrical topology.
In this simulation we used the algebraic DWG condition, and in this case we were not able to locate
a surface on which all characteristics will point inwards. Therefore,
no excision was employed in this evolution, and the eventual failure
of the code was caused by collapse of the lapse
at the axis.

While the total mass of this spacetime, $M \simeq 24.5$, is determined
by the asymptotic fall-off (\ref{asymp_to_charges}),
the mass of the ``black string'' (a black hole with the smeared horizon)
can be estimated from the average size of its horizon. The last moment of the evolution
before the code had crashed is shown in bottom panels of Fig. \ref{fig_rhoRic_t_12}. The horizon
is nearly uniform along $z$ and is located at ${\tilde r}_{AH}\sim 0.3$, which yields the
mass of $M_{BS}=0.5 R_{AH}/(G_N/{\hat L})\simeq 12 $, where $R_{AH}\sim 0.49$ is
the uncompactified average areal radius of the horizon.  This value is below
the critical mass, $M_c\simeq 14$, needed for stability of the extended solutions of this type \cite{GL}.
Therefore, at this stage the system is probably far from approaching stationarity.
Since the total available mass is higher than $M_c$ several scenarios for
subsequent dynamics can be imagined. If the black hole accretes
enough matter and increases its mass above critical, it can reach the uniform
black-string end-state. Otherwise, the evolution would probably have to proceed via forming
a localized black hole in the first place.
This black hole then may or may not accrete additional matter, and as a result either to grow and become a black string
or to settle down to a stationary solution of spherical topology.
While further investigation of the process is clearly needed, Fig. \ref{fig_rhoRic_t_12} shows
a developing progressive localization of energy-density and curvature around $(0,0)$, indicating
that formation of a localized black hole first is more probable in this specific case.

We turn now to a detailed description of the
performance of the code. Since we have implemented a free evolution scheme, we can assess the
convergence of our numerical solutions by monitoring
discrete versions of the Hamiltonian and momentum constraints, which
are defined by contracting the Einstein equations with the unit normal
vector to the $t={\rm const}$ hypersurfaces, i.e.\ $M_a \equiv
n^a(G_{ab}-T_{ab})$, where $G_{ab}$ is the Einstein tensor. One way of
doing this involves evaluation of the following $L_2$-norm of 
finite-differencing variables
\be
\label{L2_norm}
\Vert Y\Vert_{L_2}= 
{\( \frac{1}{N_r\,N_z}\sum_{i,j=1}^{N_r,N_z}|Y_{i,j}|^2 \)^{1/2}},\ee
In the next section, we compute the $L_2$-norms of the constraints at each
time step and examine their behavior as a function of various
parameters of the problem.
\subsection{Coordinate conditions}
\label{sec_results_coords}
In the weak-gravity regime where an initial pulse of matter completely disperses
to infinity we find clear advantage of the algebraic DWG conditions.
They are robust, almost do not require fine-tuning, and are extremely stable.
Additionally, as described in the next section, the constraints
remain well preserved in this case, such that only very small damping
(controlled by the parameter $\kappa$, see (\ref{cdmp})) needs to be added
to the equations.

Since at $t=0$, we assume harmonic conditions $H_a=0$,
 choosing the source functions according to ({\ref{LS_Fa}) at $t>0$ will create
 discontinuity in the temporal component of the gauge source
 function. Therefore, we
  multiply the sources $F_a$ in ({\ref{LS_Fa}) by a time-dependent function
  $(1-\exp(-t/t_0))(1+s \exp (-t/t_1))$, where $t_0, t_1$, and $s$ are
  parameters. While the first factor is used to gradually turn the
  sources on, the second factor is introduced in order to improve late
  time stability. For instance, while we found that weak data defined by $|\Phi_0| \ltsim 0.35$
 can be simulated using  $t_0 \sim 0-1$ and $ s=0$, the parameters  $t_0 \sim 0-0.2$, $s\sim 0.1-0.5$ and $t_1 \sim 10$
were required in order to maintain regularity during evolution of stronger data.
In addition, we found that long-term stability of the strong data evolutions improves when the factor $\exp(n\,S)$
is removed from the definition of the determinant
  $\gamma=(g_{11}\,g_{22}-g_{12}^2)^{1/2} \exp(n\,S)$ used in (\ref{LS_Fa}).

The simulations that employ algebraic DWG are stable for a range of the parameters $\mu_1$ and $\mu_2$. Basically,
any values of these parameters of order one can be successfully used in the collapse situations.
Nevertheless, we still find that certain choices of $\mu_{1,2}$ perform better than others
and preserve the constraints with a greater accuracy.  This is illustrated in Fig. \ref{fig_Ma_phi03_mu}
where we show norms (\ref{L2_norm}) of the constraints as a function of time. Although,
the norms decrease in all the instances, the constraints violations are the smallest for $\mu_1=\mu_2\simeq 2$.
  \begin{figure}[t!]
    \centering \noindent
    \includegraphics[width=8.0cm]{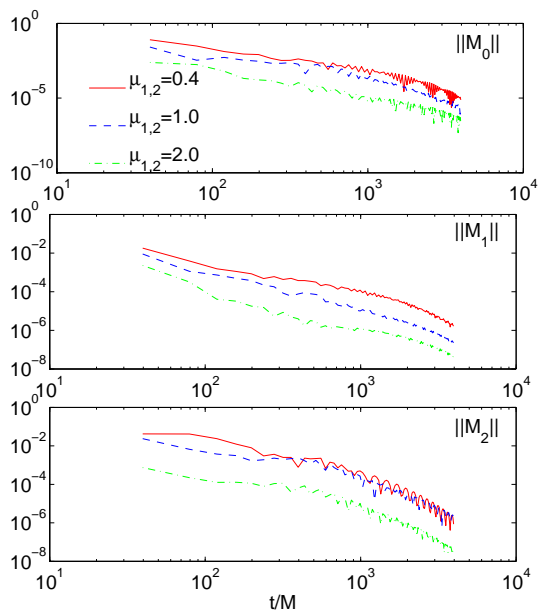}
    \caption[]{While evolutions are usually stable for any values of order one of the parameters $\mu_{1,2}$ in
      (\ref{LS_Fa}), instabilities develop if these parameters are too small or too large.
      Here we show the behavior of the norms of the constraints (\ref{L2_norm}) in the simulation
      of the date defined by $\Phi_0=0.3$ in 6D. The evolution is stable for
      $0.3\ltsim \mu_{1,2} \ltsim 5$ and the decay rate of the constraint
	violations is optimized for $\mu_{1,2} \simeq 2$. }
    \label{fig_Ma_phi03_mu}
  \end{figure}
  \begin{figure}[t!]
    \vspace{-0.6cm} \centering \noindent
    \includegraphics[width=8.0cm]{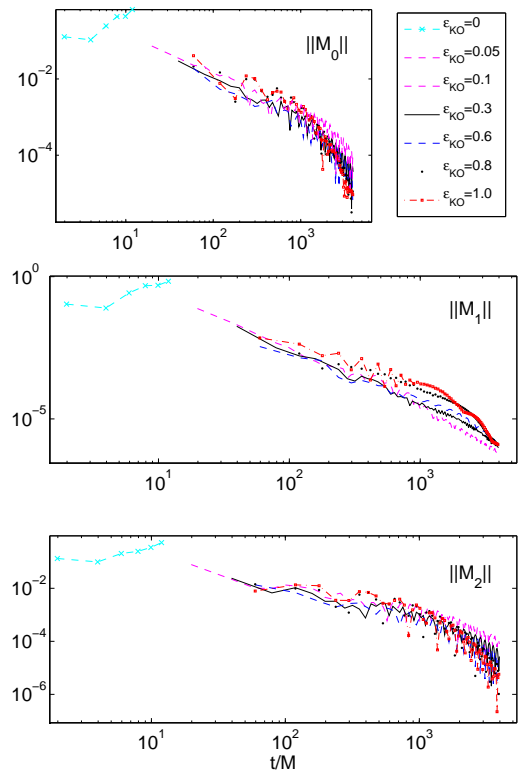}
    \caption[]{The behavior of the Hamiltonian and the momentum
      constraints, $M_a \equiv n^a(G_{ab}-T_{ab})$, as a function of the
      dissipation parameter $\eps_{KO}$ in a 6D simulation with
      $\Phi_0=0.3\,(1-i)$ and the resolution $33\times 33$.
      While for $\eps_{KO} \ltsim 0.04$ the constraints blow up and the code diverges,
	above this threshold the evolution is stable for arbitrary long periods of time, and any
      constraint violations decrease in the course of evolution.}
    \label{fig_Ma_phi03_eps}
  \end{figure}
After experimenting with the driver version of DWG condition
  (\ref{LS_driver}), we find that it performs comparably to the
  algebraic DWG in early stages of the evolution until $t\sim 10 M$. However, late-time 
behavior is fairly sensitive to the driver parameters and generically
  develops coordinate singularities on a timescale $t\sim 30-100 M$.
 A considerable constraint damping was usually required in these cases. In comparison,
a DWG-driver simulation with parameters identical to the algebraic DWG evolution shown in Fig. \ref{fig_rho_r0},
had required at least 50 times stronger damping in order not to diverge immediately,
but even then the evolution remained stable only until $t\sim 100 M$.

 The performance of the gauge
  drivers (\ref{FP_gauge}) and (\ref{FP_gauge_m}) is similar to that. It was already
  reported in \cite{SorkinChoptuik} that
  those drivers---known to perform well in Cartesian implementations
  \cite{FP3,FP2}---are considerably less stable in spherical symmetry, and here we notice the same.
  Specifically, we find that pure harmonic coordinates are useful only
  for simulating the dynamics of very weak initial data with the
  scalar amplitude $|\Phi_0| \ltsim 10^{-3}$ (and the corresponding
  mass $M \ltsim 10^{-4}$). For larger values of $|\Phi_0|$ the lapse function collapses
 in the locus of maximal matter concentration and coordinate
  singularity forms. Although we were able to delay the
  pathology by using of one the drivers
  (\ref{FP_gauge},\ref{FP_gauge_m}), in no instance was it possible to eliminate it
  completely. Generically, the coordinates in these simulations become
  singular after a time of approximately a few tens of $M$. An extensive
  fine-tuning of the driver parameters together with stronger
  constraint damping and the numerical dissipation enables one to extend somewhat
  the duration of the regular evolution. up to $t\sim 100 M$.  For
  example, a parameter setting that kept the evolution of the $|\Phi_0|=0.5$
  data regular until $t\sim 100 M$ consisted of
  $\eps_{KO}=0.34,~ \lam_C=0.1, ~\xi_1=0.6,~\xi_2=0.8,~ \xi_3=0,~
  \al_0=1, ~q=3$ and $ N_r\times N_z =41\times41$.
  \begin{figure}[t!]
    \centering \noindent
    \includegraphics[width=7.5cm]{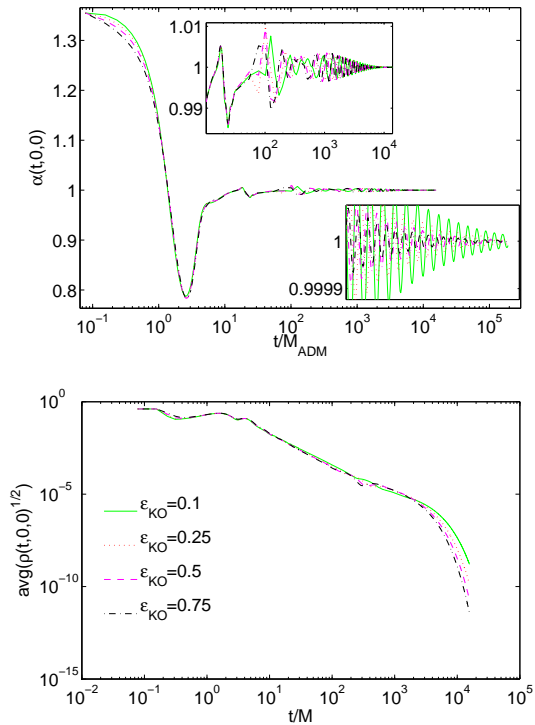}
    \caption[]{The lapse at the center and the energy-density
      averaged along the axis are shown as functions of time for various values
      of $\eps_{KO}$ in a subcritical 5D simulation, with
      $\Phi_0=0.4\,(1-i)$.  Provided we use sufficient dissipation,
      ($\eps_{KO}>0.08$ in this case), the behavior of dynamical
      variables is essentially independent of the specific $\eps_{KO}$
      in early time until $\sim 900\, M$, see the upper inset in the top
      panel. While at the late times the
      absolute values of variables in the simulations with stronger
      dissipation are generically smaller than those in simulations
      using less dissipation, the variations of the fields are
      very small at this stage in all instances, see bottom inset. }
    \label{fig_al_rho_phi04_eps}
  \end{figure}

Our preliminary experiments in supercritical regimes indicate that the algebraic DWG
still outperforms the driver conditions. While in all our simulations where a black hole
forms the coordinates eventually become singular near the excision surface, the
simulations employing algebraic DWG remain stable for longer.
It is presently unclear whether the coordinate pathology
is caused by the strong gravitational field or by the introduction of the excision surface. More experiments
are required and will be reported elsewhere.

  \subsection{Dissipation and constraint damping}
  \label{sec_eps_kappa}
  \begin{figure}[t!]
    \centering \noindent
    \includegraphics[width=7.8cm]{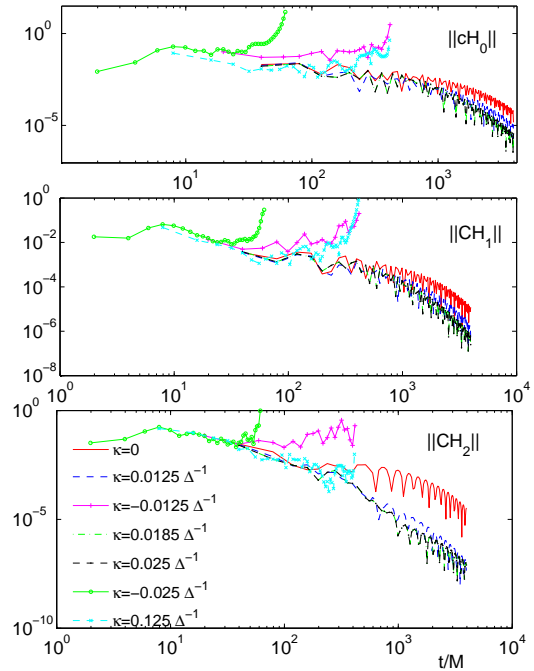}
    \caption[]{The dynamics of the norm (\ref{L2_norm}) of the GH
      constraints (\ref{Ca}) depends on the amount of the damping.
      In simulations of the initial
      data defined by $\Phi(0,r,z)=0.3\,(1-i)\exp[-(r^2+z^2)/0.25^2
      ]$ that employ algebraic DWG
      any constraint violations decrease in time,
	provided $\kappa \alt 0.0185 \Delta^{-1} $; for larger values of $\kappa$ the constraint
      violations diverge. An optimal preservation of the constraints in this case
      is achieved for $\kappa \simeq 0.05/(1+0.05\,t)$.}
    \label{fig_cHa_phi03_kp}
  \end{figure}
  \begin{figure}[t!]
    \centering \noindent
    \includegraphics[width=8cm]{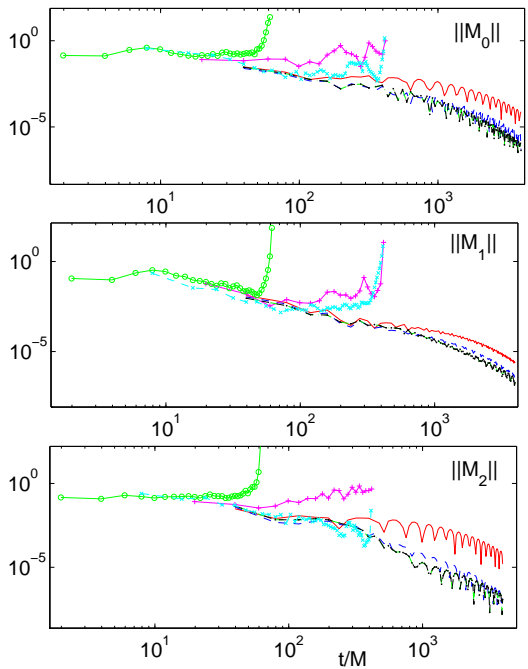}
    \caption[]{Same simulation and same notations as in
      Fig.\ref{fig_cHa_phi03_kp}.  The Hamiltonian and the momentum
      constraints are asymptotically satisfied provided a correct amount of
      damping is used.}
    \label{fig_Ma_phi03_kp}
  \end{figure}

  The explicit numerical dissipation that we add to our scheme in
  (\ref{KO_filter}) is an important ingredient affecting the long-term
  stability, and Fig.~\ref{fig_Ma_phi03_eps} illustrates this.  There
  we fix $\Phi_0=0.3\,(1-i)\exp[-(r^2+z^2)/0.25^2 ]$,
  use resolution of $33\times33$ and algebraic DWG conditions with
  $\mu_{1,2}=2$ and $q=1/2$ in a 6D simulation. We find that without the dissipation
  the constraints quickly diverge. However, for $\eps_{KO}$ above a certain minimal
 value ($\eps_{KO} >0.04$ in this case) the evolution stabilizes. The specific threshold
value increases slightly when denser grids and stronger initial
  data are used; however, taking $\eps_{KO} \simeq 0.15-0.4$ was usually a safe
  choice for the simulations described in this paper.

  In Fig. \ref{fig_al_rho_phi04_eps} we depict the lapse $\al(t,0,0)$
  and the matter energy-density $\rho(t,0,z)$, averaged along the axis,
  as functions of time for several choices of $\eps_{KO}$ in 5D
  simulations of weak scalar pulse with $\Phi_0=0.4\,(1-i)$ using algebraic DWG.
  The simulations converge for a wide
  range of the values of the dissipation parameter, provided  $\eps_{KO} \gtsim 0.08$.
  Figure \ref{fig_al_rho_phi04_eps} demonstrates that in this case the specific values of $\eps_{KO}$ have
  only a marginal effect on the early dynamics, until approximately $
  1000\,M$.  However, after that time the variables computed in
  the simulations using unequal dissipation parameters begin to differ,
  and stronger dissipation generically implies smaller late-time
  amplitudes. The absolute values of the amplitudes are usually
  small at this stage (typically below $\sim 10^{-4}$).

  Another factor influencing stability is the constraint damping term
  (\ref{cdmp}), which we add to the Einstein equations in
  (\ref{Eqs_constrdamp}).  We find that quite generically the long-term
  stability improves when the damping of the constraints vanishes
  in the asymptotic regions. For this reason we multiply $\kappa$ by a
  factor $(R_0/R)$ in the regions where the areal radius satisfies
  $R>R_0$ for some large $R_0$ (typically $R_0\sim 20$) in order to
  gradually turn the damping off.

  Figures \ref{fig_cHa_phi03_kp} and \ref{fig_Ma_phi03_kp} illustrate the
  effect of the damping in the case
  $\Phi(0,r,z)=0.3\,(1-i)\exp[-(r^2+z^2)/0.25^2 ],
  ~\eps_{KO}=0.5$ in 6D simulations that use algebraic DWG with
  $\mu_{1,2}=1$ and $q=1/2$, and the resolution of $33\times33$.  In this
  configuration the evolution is stable for rather small values of
  $\kappa$ including that of $\kappa=0$; however, the quickest asymptotic
  decrease of the constraint violations is achieved for
  $\kappa=0.05/(1+0.05\,t)$. For the values of $\kappa$ greater than a
  certain value---$\kappa > 0.25$ in this specific
  simulation---the code diverges.  The time-dependent factor
  $1/(1+0.05\,t)$ is less crucial in shorter subcritical simulations,
  but it helps to improve stability on the long time scales of $t\gtsim
  2000\, M$.

  Interestingly, typical values of the damping parameter $\kappa$ in
  algebraic DWG evolution are small compared to the
  inverse of any typical length scales of the problem (set e.g. by the
  initial width of the scalar pulse, $\Delta$, by the size of the
  KK circle or by the size of the black hole).  Moreover, certain weaker initial data can even be
  simulated without the damping at all.  This is in sharp contrast
  to the typical values of the damping parameter, $\kappa~\simeq
  1/\Delta$, required in simulations that employ the driver gauge conditions
  (\ref{LS_driver}), (\ref{FP_gauge}), or (\ref{FP_gauge_m}).
  \subsection{Convergence}
  \label{sec_convergence}

  \begin{figure}[!t]
    \centering \noindent
    \includegraphics[width=8.5cm,height=9.0cm]{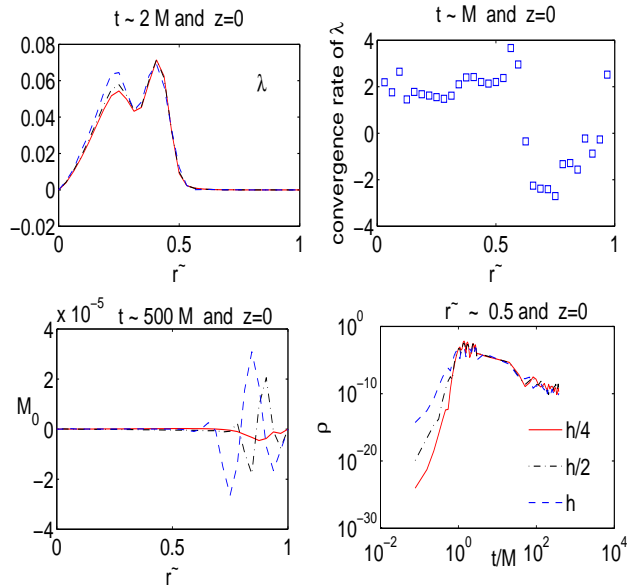}
    \caption[]{A convergence in a 5D algebraic DWG evolution of
      $\Phi_0=0.4\,(1-i)$, with $h=1/32$. Top panels indicate that
      in the regions where a non-negligible amount of
      matter is present the convergence rate
      (\ref{logc2c1}) of the metric function $\lambda$ is essentially
      $p \sim 2$. Other functions, such as energy
      density, depicted in the bottom right panel, have similar
      convergence trends.  The bottom left panel illustrates that the
      constraint violations consistently decrease with increasing resolution.}
    \label{fig_convergence}
  \end{figure}
  \begin{figure}[t!]
    \centering \noindent
    \includegraphics[width=8cm]{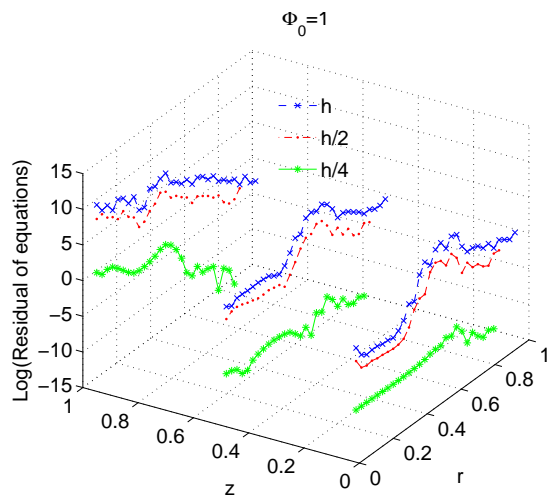}
    \caption[]{The total residual (\ref{res}) of the system of our FDA equations at
      the hypersurfaces $z=0, z=1/2$ and $z=1$ for 3 resolutions, with
      the coarsest one $h=1/32$. The residual quickly
      decreases as a function of the numerical resolution. }
    \label{fig_L2eqs_phi1}
  \end{figure}

  One of the crucial tests of numerical FDA schemes, such as one
  that we use here, involves the investigation of the convergence of
  the generated numerical solutions as a function of resolution.  We
  perform convergence tests based on the assumption \cite{Richardson}
  that for any of the unknown functions, $Y(t,r,z)$, appearing in our
  system, the corresponding discrete quantity, $Y_h(t,r,z)$ in the
  limit $h\to0$ admits an asymptotic expansion of the form
  \be
  \label{richardson}
  Y_h(t,r,z) = Y(t,r,z) + h^p e_p(t,r,z) + \cdots \ee
  where $h$ is the spatial mesh size, $e_p(t,r,z)$ is an
  $h$-independent error function, and $p$ is an integer that defines the order of
  convergence of the scheme.  We consider sequences of three
  calculations performed with identical initial conditions, but with
  varying resolutions, $h$, $h/2$ and $h/4$, and compute
  \be
  \label{logc2c1}
  \log_2\(\frac{Y_{h}-Y_{h/2}}{Y_{h/2}-Y_{h/4}}\) \approx p, \ee
  for a grid function $Y$.

  We fix $\Phi_0=0.4\,(1-i)$ and use algebraic DWG evolution in 5D
  with the parameters $\mu_{1,2}=3, q=1/2,\kappa=0.07, \eps=0.125$,
  and show in Fig.~\ref{fig_convergence} several plots illustrating
  the convergence.  The top left panel shows the radial dependence of the
  metric function $\lambda$ along $z=0$ at $t\sim 2\,M$
  obtained in simulations using 3 different resolutions, $h, h/2$ and
  $h/4$, with $h=1/32$. The decreasing differences between solutions obtained
using increasing resolutions indicate convergence.  The top
  right panel in Fig.~\ref{fig_convergence} shows that the convergence rate
  (\ref{logc2c1}) is mostly quadratic, except in the
asymptotic region where the amplitude of the field is small and the computation is unreliable.
The bottom left panel depicts the Hamiltonian constraint at
  a late moment of the evolution:  violations of the constraint are
  small and decrease further when the grid is refined. Finally, the bottom
  right panel illustrates the time variation of the logarithm of the
  matter energy-density at $(\tilde{r},z)\sim(0.5,0)$. Again, a
 convergence is evident and is compatible with $p \gtsim 1.5$ in the
  regions where a non-negligible amount of matter is present.

  An additional measure of accuracy of the scheme can be obtained by
  monitoring the total normalized residual of our
equations that can be defined as
  \be
  \label{res} { \cal R} =
 \(\frac{\sum_{i,j=1}^{N_r,N_z} \sum_{Y} |{{\cal
        R}_Y}_{i,j}|^2}{{N_r\,N_z}\,\sum_Y 1}\)^{1/2} \ee
  where ${{\cal R}_Y}_{i,j}$ is the residual of the FDA equation
  governing the variable $Y$ at a grid-point $(i,j)$.  We show the
  logarithm of the residual in Fig. \ref{fig_L2eqs_phi1} for three
  resolutions $h, h/2$ and $h/4$, with $h=1/32$ in the 6D simulation
  using algebraic DWG and the initial scalar pulse amplitude
  $\Phi_0=1.0 \exp(-(r^2+z^2)/0.125)$. Evidently, the residual quickly decreases
  as a function of the numerical resolution, again indicating convergence of the scheme.

  \subsection{Conclusion}
  \label{sec_conclusion}

  We have described a generalized harmonic formulation of the Einstein
  equations in axial symmetry in $D$-dimensions and constructed the
  first numerical code based on it. We chose the coordinates in which the
  background symmetries are explicit. This, however, resulted in a
  coordinate singularity on the axis, $r=0$.  While at the continuum
  level the equations of motion ensure regularity of a solution on the
  axis, extra care must be exercised so that this remains true in
  discrete numerical calculations.  We have devised a regularization
  procedure that achieves that, while preserving the hyperbolicity of
  the evolution system.  In our implementation we integrate the full
  $D$-dimensional equations and find that this approach is smoother at
  the axis and is generically more stable compared to the approach
  that solves the 2+1 equations, obtained by a dimensional
  reduction on the symmetry.

  We expect that our new code will enable systematic investigation
  of many problems of interest in axisymmetry.  As a first application
  we tested the performance of the code in the context of fully
  nonlinear gravitational collapse of a complex, self-interacting
  scalar field propagating in a $D$-dimensional Kaluza-Klein spacetime. We assumed spherical
  symmetry in the $(D-1)$-dimensional noncompact portion of the spacetime,
   which effectively reduced the problem to 2+1. The
  scenarios that we have considered ranged from the dispersion of
  dilute pulses to the collapse of strongly gravitating pulses that
  lead to black hole formation. One of the aspects of our code was the
  use of radial compactification which, in conjunction with sufficient
  Kreiss-Oliger--type dissipation, provided a viable alternative to the
  truncation of the spatial domain and the use of approximate outer
  boundary conditions. Another ingredient of our algorithm that in some regimes was
  vital for long-term stability was the addition of
  constraint-damping terms to the evolution equations.

  We described several strategies to fixing the coordinate freedom that are
  compatible with the GH approach and experimented with those. Our
  studies of evolutions using damped wave gauge that was enforced
  algebraically indicate robustness of this choice, its weak
  dependence on parameter settings, and stability. On the other hand,
  our experiments with various drivers---described in detail in Sec.
  \ref{sec_coord_condition}---reveal that in the case of symmetries
  these drivers are considerably less effective relative to the $3+1$
  simulations that use Cartesian coordinates, a conclusion similar to that
  drawn in \cite{SorkinChoptuik}. However, it would be interesting
  to examine performance of the drivers in additional axisymmetric
situations, other than collapse.

  One can consider several ways to improve the code.
  Specifically, it seems natural to use hyperboloidal slicing, similar
  to one suggested in \cite{anil}, instead of the spatial compactification
  that we presently employ. We expect this will allow calculating
  asymptotic quantities and emitted gravity wave with a greater accuracy,
  and will help to improve late-time stability of the evolution, that
  is adversely affected by the unphysical radiation caused by the loss of numerical 
  resolution near the outer boundary. 
  One possible extension of our code which will
  broaden significantly the spectrum of
  possible axisymmetric configurations accessible with it, would be an
  addition of rotation.
  In addition, coupling the gravity dynamics to evolution
  of more general matter, such as fluid, will enable studying certain
  situations relevant in astrophysics.

  Although the main purpose of this work was to describe the
  code and test its performance in a non-trivial highly dynamical
  setting, our preliminary study of collapse in a Kaluza-Klein
  background indicates rich and distinctive phenomenology,
  deserving further investigation. Among interesting questions, which
  will be addressed elsewhere, is determining
  what classes of initial data lead to formation of black holes of specific topology
  and constructing a phase diagram of the solutions (see \cite{BH-BS} for some predictions
  concerning the diagram). In addition, a detailed investigation of the situation near
  threshold for black hole formation\footnote{This is exactly where we expect the
    Adaptive Mesh Refinement (AMR) feature provided by the pamr/amrd
    infrastructure, will be crucial.}
and classification of possible outcomes is necessary.
  In particular, in this regime the fields
  typically oscillate, creating several outgoing shells.
  We found that the matter distributes anisotropically within the
  shells, forming separate bounded systems, and that
curvature singularities can develop inside those.
 We expect that the effect will be more pronounced in higher dimensions
 where many more shells can
  form near threshold \cite{SorkinOren}, and
  since the gravitational field of an isolated system
  is increasingly localized in higher dimensions, the shells have
  a greater chance to separately become bounded systems capable of collapsing
  and forming black holes.
  It is worth exploring to which extent the
  expectations are true. However, such a study will require an improved horizon-finder that
will be able to locate several moving horizons, and we are working in this direction.

  Finally, it would be very interesting to compute the
  detailed gravitational-wave signal emitted in various regimes.
  While from the perspective of a four-dimensional observer all our
  solutions appear spherical that are not expected to emit any
  gravitational radiation, the waves are certainly produced and carry energy away.
  The ``missing'' energy, as it would be seen in 4D, will then provide a circumferential evidence in favour of the extra-dimensions.
  In addition, the spectrum of the gravitational waves must contain frequencies associated with the
  length scale of the extra dimensions. Therefore, by measuring the signal
  directly one can, in principle, probe the dimension and the topology of
  the spacetime.

 \begin{acknowledgments} I would like to thank Luciano Rezzolla and  Badri Krishnan for
    useful discussions, Matt Choptuik for collaboration on related projects,
    and Choptuik and Aaryn Tonita for valuable comments on the manuscript.
    The computations were performed on the Damiana
    cluster of AEI.
  \end{acknowledgments}
  \appendix
  \section{Dimensionally reduced equations and boundary conditions}
  \label{sec_KK}
  Here we describe the approach that uses the Kaluza-Klein reduction
  of the $D$ dimensional equations on the $O(D-2)$ symmetry and
  integrates the resulting three-dimensional equations.

  The most general $D$-dimensional line element that is invariant
  under action of the group of rotational symmetries $O(D-2)$ can be
  written as
  \bea
  \label{SO_D-2_metric_KK}
  ds^2&=& e^{2\,\al \,\h S} ds_3^2 + e^{2\,\bt\,\h S} d\Omega_n^2 \non
  &=&e^{2\,\al \,\h S} g_{ab}^{(3)} dx^a dx^b + e^{2\,\bt\,\h S}
  d\Omega_n^2 , \eea
  where the metric components are functions of three-dimensional
  coordinates $x^a$ alone, and $\al, \bt$ are constants, chosen for
  convenience below.

  We define the conformally rescaled metric $ \tilde{g}_{ab}^{(3)}
  \equiv e^{2\,\al \,\h S} g_{ab}^{(3)}$ and write the dimensional
  reduction of the Einstein-Hilbert Lagrangian as
  \bea \label{Sreduce1} \cL_{EH} &\equiv &\sqrt{-g_D} R_D = \non &=&
  \sqrt{-\tl g_3 } e^{n\, \bt\, \h S} \Big[ \tl R_3 + n
  (n-1)\,e^{-2\,\bt\,\h S} -\non &&-2\,n\,\bt\,\tl\Box\,\h S
  -n(n+1)\,\bt^2\, (\tl \pa \,\h S)^2 \Big], \eea
  where derivatives are computed using $ {\tl g^{(3) ab}}$: $\tl\Box
  \,\h S \equiv (-\tl g_3)^{-1/2} \pa_a \left( {\tl g^{(3) ab}}\,(-\tl
    g_3)^{1/2}\, \pa_b \, \h S \right) $, and $\left( \tl\pa \h
    S\right)^2 \equiv {\tl g^{(3) ab}}\pa_a \h S \pa_b \h S$.

  Substituting the non-tilded metric and using the relations between
  conformally related quantities \cite{Wald},
  \bea
  \label{Weyl1}
  \tl R_3&=& e^{-2\,\al\,\h S} \[R_3-4\,\al\, \Box \,\h S - 2\,\al^2
  \,(\pa \h S)^2\],\non \tl \Box \,\h S &=& e^{-2\,\al\,\h S} \[ \al
  \,(\pa \,\h S)^2 + \Box \,\h S\], \non \left( \tl\pa \h S\right)^2
  &=& e^{-2\,\al\,\h S} (\pa\h S)^2, \eea
  where the derivatives in the right-hand-side are now computed with
  the non-tilded metric, we arrive at
  \bea
  \label{Sreduce2}
  \sqrt{-g_D} R_D& =& \sqrt{- g_3 } e^{(\al+n\, \bt) \h S} \Big[ R_3
  -(4 \al+ 2\,n\,\bt) \Box \h S \non &-& \(2\,\al^2 +
  2\,n\,\al\,\bt+n(n+1)\bt^2\) (\pa\h S)^2 + \non && n
  (n-1)\,e^{-2\,\bt\,\h S +2\,\al\,\h S} \Big].  \eea
  By choosing $\al=-n\,\bt$, we convert the lower-dimensional action
  into Einstein-Hilbert form, $\cL = \sqrt{-g_3} R_3 $. Since in this
  case $ \Box \h S$ is multiplied by a constant factor, it does not
  contribute to the equations of motion and can be omitted.  In
  addition, fixing $ \bt = [n\,(n+1)]^{-1/2}$ ensures the canonical
  normalization of the kinetic term of the scalar $\h S$. After these
  manipulations the Lagrangian becomes
  \be \label{Sreduce} \cL_{EH}= \sqrt{- g_3 }\[ R_3 - ( \pa \h S)^2+n
  (n-1)\,e^{-2\,c_n \h S } \], \ee
  where we have defined $c_n \equiv \sqrt{(n+1)/n}$.  This Lagrangian
  describes three-dimensional gravity coupled to the self
  interacting\footnote{Note that the potential term of the scalar $\h
    S$ is proportional to the curvature of the $n$-sphere, and hence
    the scalar $\h S$ is massless in axisymmetry in 4D.}  scalar field
  $S$.  The components of the original $D$-dimensional metric
  $g_{\mu\nu}$ (\ref{SO_D-2_metric_KK}) are given in terms of the
  lower-dimensional fields as
  \bea
  \label{gcomponentsSO_D-2}
  g^{(D)}_{ab}&=& e^{-(2/c_n) \, \h S}\, g^{(3)}_{ab}, \non
  g^{(D)}_{\Omega\Omega}&=& \exp\[\frac{2}{\sqrt{n(n+1)}} \h S\]\,
  g_{\Omega}.  \eea

  A reduction of the matter Lagrangian of our model (\ref{action})
  takes the form
  \be
  \label{LPhi}
  \cL_{\Phi}= \half \sqrt{-g_3} \[ | \pa \Phi|^2 +
  2\,V(|\Phi|)\,e^{-(2/c_n) \, \h S} \], \ee
  yielding the total action of the system,
  \bea
  \label{L1}
  S &=& S_{EH}+ S_{\Phi} = \non &-&\frac{1}{2} \int\sqrt{- g_3 }\Big[
  R_3 -|\pa \Phi|^2 - ( \pa \h S)^2 + \non &+&n (n-1)\,e^{-2\,c_n \h S
  } -2\, V(|\Phi|)\,e^{-(2/c_n) \, \h S} \Big], \eea
  that describes two interacting scalar fields minimally coupled to
  gravity in three dimensions.  Varying the action with respect to the
  3-metric and the fields one obtains the equations of motion,
  \bea
  \label{Eqs_KK}
  R_{ab} &=& \bar{T}_{ab} \equiv T_{ab}-g_{ab}\, T = \non && =\pa_{(a}
  \Phi \, \pa_{b)} \Phi^* + \pa_a \h S \,\pa_b \h S -\non && -g_{ab}
  \(-2\,V\,e^{-(2/c_n) \, \h S}+n(n-1)e^{-2\,c_n \h S}\), \non \Box \h
  S & -& c_n\,n(n-1) e^{-2\,c_n \, \h S } +
  \frac{2}{c_n}\,V\,e^{-(2/c_n) \, \h S} =0 , \non \Box \Phi & -&
  \frac{\pa V }{\pa \Phi^*}e^{-(2/c_n) \, \h S} =0.  \eea
  Next, we bring the Einstein equations in this system into the
  GH form using the transformations analogous to (\ref{Ca}),
  (\ref{eqH}) but applied to the 3-metric, and add a constraint
  damping term similar to (\ref{cdmp}). The resulting hyperbolic
  system is evolved in time.

  We are interested in asymptotically Minkowski times a circle,
  $\mathbb{R}^{D-2,1}\times S^1$ solutions of (\ref{Eqs_KK}),
  satisfying $g^{(D)}_{ab} \rightarrow \eta_{ab}$,
  $g^{(D)}_{\Omega\Omega} \rightarrow r^2$, and $\Phi \rightarrow 0$.
  Using (\ref{gcomponentsSO_D-2}) we find that asymptotic boundary
  conditions obeyed by the reduced fields in this case are
  $g^{(3)}_{ab}\rightarrow \eta_{ab} r^{2\,n}$ and $\h S \rightarrow
  \sqrt{n(n+1)} \log(r)$.  Since it is difficult to handle blowing-up
  conditions of this sort in numerical implementations, we redefine
  our variables by factoring out this singular behavior: $g_{ab}
  \rightarrow g_{ab} r^{2\,n}$ and $S = \hat S + \sqrt{n(n+1)}
  \log(r)$. However, as a result of this transformation, the radial
  component of the GH source functions defined in (\ref{GH_coords}),
  acquires a term singular at the axis $H_1=n/r+\dots$, where ellipses
  designate regular at $r=0$ terms. This behavior is similar to what
  happens in the unreduced system and in analogy to that case we
  regularize the source functions by subtracting off this singular
  flat-background contribution, see Sec. \ref{sec_axis}. The gauge
  conditions are then applied to the regularized source functions.

  Regularity conditions at the center $r=0$ are again analogous to
  those in the unreduced case. Specifically $g^{(3)}_{00},
  g^{(3)}_{11}, g^{(3)}_{02}, g^{(3)}_{22}$ and $S$ are even functions
  in $r$ as $r\to0$, while $g^{(3)}_{01}$ and $g^{(3)}_{12}$ are odd.
  Moreover, requiring the absence of conical singularity at $r=0$ places
  an additional condition that $g^{(3)}_{11} -\exp(-c_n\, S) ={\cal
    O}(r^2)$ as $r\to0$.  Since it is desirable to have a number of
  boundary conditions matching the number of dynamical variables, one
  might consider a regularization similar to (\ref{lambda_var}) that
  achieves this by defining $\lam = ( {g_{11}} -e^{c_n\, S})/r$ that
  behaves as ${\lam}\sim {\cal O}(r)$ near $r=0$,  one than eliminates $S$ from the
  scheme and uses $\lam$ as a fundamental variable instead.  However, a
  closer examination of the equation governing $\lam$ reveals that the
  regularization does not work in this case because the equation is
  not automatically regular at $r=0$ as it was in the unreduced
  approach. Rather the equation contains term proportional to $1/r$,
   which is regular only if the
  constraints are explicitly satisfied. Namely, not only the number of boundary
  conditions exceeds that of the dynamical fields---as it was before the
  regularization---but now the extra condition is also algebraically more
  complicated. Hence, instead of introducing $\lam$ we opted to
  implement a more straightforward regularization method that
  maintains $S$ as a fundamental dynamical variable and uses
  analytical Taylor-series expansion to compute its value near the
  axis; see \cite{SorkinChoptuik} for more details.

  A comparison of the dimensionally reduced approach and the full
  $D$-dimensional method described in the main text, shows that both
  methods perform comparably in the weak-field regime when $M \ltsim 0.5$.
  However, for stronger initial data the dimensionally-reduced
  approach is considerably less stable and prone to developing
  coordinate singularities, regardless of the gauge conditions that we use.
  We do not fully understand the reason for
  this, a further investigation is required, and we hope to report on
  this puzzling issue elsewhere.



\begin{thebibliography}{99}
\bibitem{FP2} F.~Pretorius, 
  Phys.\ Rev.\ Lett.\ {\bf 95}, 121101 (2005)
\bibitem{FP1}
  F.~Pretorius, 
  Class.\ Quant.\ Grav.\ {\bf 22}, 425 (2005)

\bibitem{FP3} F.~Pretorius, 
  Class.\ Quant.\ Grav.\ {\bf 23}, S529 (2006)

\bibitem{RIT} M.~Campanelli, C.~O.~Lousto, P.~Marronetti and
  Y.~Zlochower,
  Phys.\ Rev.\ Lett.\ {\bf 96}, 111101 (2006)
\bibitem{nasa} J.~G.~Baker, J.~Centrella, D.~I.~Choi, M.~Koppitz and
  J.~van Meter,
  Phys.\ Rev.\ Lett.\ {\bf 96}, 111102 (2006)
\bibitem{AEI} M.~Koppitz, D.~Pollney, C.~Reisswig, L.~Rezzolla,
  J.~Thornburg, P.~Diener and E.~Schnetter,
  Phys.\ Rev.\ Lett.\ {\bf 99}, 041102 (2007)
\bibitem{CaltechCornell} M.~Boyle {\it et al.},
  Phys.\ Rev.\ D {\bf 76}, 124038 (2007)

\bibitem{Garfinkle_axi}
  D.~Garfinkle and G.~C.~Duncan,
  Phys.\ Rev.\  D {\bf 63}, 044011 (2001)

\bibitem{Choptuik_axi} M.~W.~Choptuik, E.~W.~Hirschmann,
  S.~L.~Liebling and F.~Pretorius,
  Class.\ Quant.\ Grav.\ {\bf 20}, 1857 (2003)

\bibitem{Rinne_axi_1} O.~Rinne,
  Class.\ Quant.\ Grav.\ {\bf 25}, 135009 (2008)
\bibitem{Rinne_axi_2} O.~Rinne,
  Class.\ Quant.\ Grav. {\bf 27}, 035014 (2010).

\bibitem{cartoon_method} M.~Alcubierre, S.~Brandt, B.~Bruegmann,
  D.~Holz, E.~Seidel, R.~Takahashi and J.~Thornburg,
  Int.\ J.\ Mod.\ Phys.\ D {\bf 10}, 273 (2001)


  \bibitem{BrillWaves} A.~M.~Abrahams, K.~R.~Heiderich, S.~L.~Shapiro
  and S.~A.~Teukolsky,
  Phys.\ Rev.\ D {\bf 46}, 2452 (1992).
  D.~Garfinkle and G.~C.~Duncan,
  Phys.\ Rev.\ D {\bf 63}, 044011 (2001),~

\bibitem{AbrahamsEvans} A.~M.~Abrahams and C.~R.~Evans,
  Phys.\ Rev.\ Lett.\ {\bf 70}, 2980 (1993).

\bibitem{crit_collapse_axi} M.~W.~Choptuik, E.~W.~Hirschmann,
  S.~L.~Liebling and F.~Pretorius,
  Phys.\ Rev.\ D {\bf 68}, 044007 (2003)

\bibitem{BlackRing} R.~Emparan and H.~S.~Reall, 
  Phys.\ Rev.\ Lett.\ {\bf 88}, 101101 (2002)

\bibitem{MP} R.~C.~Myers and M.~J.~Perry, 
  Annals Phys.\ {\bf 172}, 304 (1986).

\bibitem{GL} R.~Gregory and R.~Laflamme, 
  Phys.\ Rev.\ Lett.\ {\bf 70}, 2837 (1993)

\bibitem{HM} G.~T.~Horowitz and K.~Maeda,
  Phys.\ Rev.\ Lett.\ {\bf 87}, 131301 (2001)

\bibitem{Choptuik_BS} M.~W.~Choptuik, L.~Lehner, I.~Olabarrieta,
  R.~Petryk, F.~Pretorius and
  H.~Villegas, 
  Phys.\ Rev.\ D {\bf 68}, 044001 (2003)

\bibitem{Friedrich85} H.
  Friedrich. 
  Commun.  Math. Phys., 100:525-543, 1985

\bibitem{Friedrich96} H.~Friedrich, 
  Class.\ Quant.\ Grav.\ {\bf 13}, 1451 (1996).

\bibitem{Garfinkle2001}
  D.~Garfinkle,
  Phys.\ Rev.\ D {\bf 65}, 044029 (2002)

\bibitem{Lindblom_etal} L.~Lindblom, M.~A.~Scheel, L.~E.~Kidder,
  R.~Owen and O.~Rinne, 
  Class.\ Quant.\ Grav.\ {\bf 23}, S447 (2006)

\bibitem{LS} L.~Lindblom and B.~Szilagyi,
  arXiv:0904.4873.

\bibitem{SorkinChoptuik} E.~Sorkin and M.~W.~Choptuik, 
  to appear in Gen.~Rel.~Grav.~ (2010)
  arXiv:0908.2500 [gr-qc].



\bibitem{KST} L.~E.~Kidder, M.~A.~Scheel and S.~A.~Teukolsky,
  Phys.\ Rev.\ D {\bf 64}, 064017 (2001)

\bibitem{gm_sys} O.~Brodbeck, S.~Frittelli, P.~Hubner and O.~A.~Reula,
  J.\ Math.\ Phys.\ {\bf 40}, 909 (1999)

\bibitem{Gundlachetal} C.~Gundlach, J.~M.~Martin-Garcia, G.~Calabrese
  and I.~Hinder, 
  Class.\ Quant.\ Grav.\ {\bf 22}, 3767 (2005)



\bibitem{lambda_ref} A.~Arbona and C.~Bona, 
  Comput.\ Phys.\ Commun.\ {\bf 118}, 229 (1999).
  M.~Alcubierre and J.~A.~Gonzalez, 
  Comput.\ Phys.\ Commun.\ {\bf 167}, 76 (2005)

\bibitem{SorkinOren} E.~Sorkin and Y.~Oren,
  Phys.\ Rev.\ D {\bf 71}, 124005 (2005).

\bibitem{PretoriusChoptuik_coll} M.~W.~Choptuik and F.~Pretorius,
  arXiv:0908.1780 [gr-qc].


\bibitem{gamma_dr} J.~R.~van Meter, J.~G.~Baker, M.~Koppitz and
  D.~I.~Choi, 
  Phys.\ Rev.\ D {\bf 73}, 124011 (2006)


\bibitem{Scheel_etal} M.~A.~Scheel, M.~Boyle, T.~Chu, L.~E.~Kidder,
  K.~D.~Matthews and H.~P.~Pfeiffer, 
  arXiv:0810.1767 [gr-qc].
\bibitem{Lindblom_etal_gauge} L.~Lindblom, K.~D.~Matthews, O.~Rinne
  and M.~A.~Scheel, 
  arXiv:0711.2084 [gr-qc].


\bibitem{KolSorkinPiran1} B.~Kol, E.~Sorkin and T.~Piran, 
  Phys.\ Rev.\ D {\bf 69}, 064031 (2004)
\bibitem{HO1} T.~Harmark and N.~A.~Obers,
  Class.\ Quant.\ Grav.\ {\bf 21}, 1709 (2004)



\bibitem{KO} H. Kreiss and J. Oliger, 
  GARP Report No. 10, 1973


\bibitem{York} J.W. York, Jr., in Sources of Gravitational Radiation,
  ed. L. Smarr, Seattle, Cambridge University Press (1979).

\bibitem{SorkinPiran} E.~Sorkin and T.~Piran,
  Phys.\ Rev.\ Lett.\ {\bf 90}, 171301 (2003).

\bibitem{thornburg_excision} J Thornburg, Class.\ Quant.\ Grav.\ {\bf 4}, 1119, (1987)

\bibitem{pamr_amrd} Parallel Adaptive Mesh Refinement (PAMR) and
  Adaptive Mesh Refinement Driver (AMRD),
  \texttt{http://laplace.phas.ubc.ca/Group/Software.html}.



\bibitem{Richardson}
  L.~F.~Richardson, 
  Phil. Trans.\ Roy.\ Soc. A {\bf 210}, 307 (1911)


\bibitem{anil} A.~Zenginoglu,
  Class.\ Quant.\ Grav.\ {\bf 25}, 195025 (2008)

\bibitem{BH-BS} B.~Kol,
  Phys.\ Rept.\ {\bf 422}, 119 (2006),
  T.~Harmark and N.~A.~Obers,
  arXiv:hep-th/0503020.
\bibitem{Wald} R. Wald, General Relativity, University of Chicago Press, Chicago, 1984
\end{thebibliography}
\end{document}